\documentclass[twocolumn]{emulateapj}

\usepackage{natbib}
\usepackage{color}
\begin{document}
\title{Parameterizing the Supernova Engine and its Effect on Remnants
  and Basic Yields}

\author{Chris L. Fryer\altaffilmark{1,2,3}, Sydney
  Andrews\altaffilmark{1,4}, Wesley Even\altaffilmark{1,5}, Alex
  Heger\altaffilmark{6,7}, Samar Safi-Harb\altaffilmark{8}}

\altaffiltext{1}{CCS Division, Los Alamos National Laboratory, 
Los Alamos, NM 87545}
\altaffiltext{2}{Department of Astronomy, The University of Arizona,
Tucson, AZ 85721} 
\altaffiltext{3}{Department of Physics and Astronomy, The University of New Mexico,
Albuquerque, NM 87131}
\altaffiltext{4}{Physics Dept., New Mexico Tech, Socorro, NM 87801}
\altaffiltext{5}{Department of Physical Sciences, Southern Utah University, Cedar City, UT 84120}
\altaffiltext{6}{Monash Center for Astrophysics, School of Physics and Astronomy, Monash University, Victoria, 3800}
\altaffiltext{7}{School of Physics and Astronomy, University of Minnesota, Minneapolis, MN 55455} 
\altaffiltext{8}{Department of Physics and Astronomy, University of Manitoba, Winnipeg, MB R3T 2N2, Canada}

\begin{abstract}

Core-collapse supernova science is now entering an era where engine
models are beginning to make both qualitative and quantitative
predictions.  Although the evidence in support of the convective
engine for core-collapse supernova continues to grow, it is difficult
to place quantitative constraints on this engine.  Some studies have
made specific predictions for the remnant distribution from the
convective engine, but the results differ between different groups.
Here we use a broad parameterization for the supernova engine to 
understand the differences between distinct studies.  With this
broader set of models, we place error bars on the remnant mass and
basic yields from the uncertainties in the explosive engine.  We find
that, even with only 3 progenitors and a narrow range of explosion
energies, we can produce a wide range of remnant masses and
nucleosynthetic yields.

\end{abstract}


\section{Introduction}

A core-collapse supernova is produced when the core of a massive star
collapses under its own weight, forming a neutron star.  The
gravitational potential energy released in this collapse is
$2-3\times10^{53}$\,erg.  Extracting 1\% of this energy has been the
focus of core-collapse engine astronomers for the past 80
years~\citep{burrows13}.  The current favored engine behind supernova
invokes convection above the neutron star formed in the collapse.
When the collapse of the core is stopped by nuclear forces and neutron
degeneracy pressure, it sends a bounce shock through the star that
quickly stalls.  The region between the proto-neutron star and the
stalled shock is convectively unstable~\citep{herant94}.  An
increasing number of simulations have produced explosions under this
paradigm, and it has slowly gained traction and become the leading
theory model for the supernova
engine~\citep{fryer07,takiwaki14,melson15,lentz15,burrows16}.

Additionally, the observational support for the convective supernova
engine continues to grow~(for a review, see Fryer et al. 2017,
submitted).  One of the strongest pieces of evidence demonstrating
that at least some supernovae are powered by this convective engine is
the recent observations of the $^{44}$Ti distribution in the
Cassiopeia A remnant~\citep{grefenstette14,grefenstette17}.  Unlike
other elements observed in supernova remnants that emit only when they
are shock heated, the NuSTAR observations detect decay lines from
$^{44}$Ti, observing all of the $^{44}$Ti, including unshocked
material.  Since $^{44}$Ti is produced in the innermost ejecta, it
provides an ideal probe of the engine asymmetries.  The multi-mode but
not bimodal structure of the $^{44}$Ti distribution observed by NuSTAR
not only requires engine asymmetries, but rules out the jet/magnetar
engines for the supernova that produced the Cassiopeia A
remnant~\citep{grefenstette14}.  With these observations, the
convective engine is now the clear leading theory for supernova
explosions.

Many details of this convective engine remain unknown.  For example,
the time it takes to drive an explosion varies with simulations.
Similarly, the explosion energy is sensitive to the simulation
details.  The uncertainties lead to a range of predictions for the
remnant masses and yields of a given progenitor.  Accurate remnant
masses and yield measurements may provide clues to the nature of this
convective engine, but to understand these clues, we must first
understand the range of results from this engine paradigm.

In this paper, we develop a broadly parameterized model to study more
fully the possible solutions from the convection-enhanced supernova
engine.  To capture the convective engine more accurately, we
implement a 3-part parameterization for the energy injection: power,
duration and the extent of the energy injection region.  With these
models, we produce a range of explosion energies, compact remnant
masses, and nucleosynthetic yields.  In Section~\ref{sec:remnant}, we
describe details of past remnant mass calculations including aspects
of the explosion that set the mass of the compact remnant.
Section~\ref{sec:simulations} describes our progenitors and our method
to implement explosions.  In Section~\ref{sec:results}, we review the
results of our simulation suite, studying both the remnant mass and
key yields typically observed in supernova remnants.  Without
constraints, a wide range of results are possible.  By limiting the
allowed explosion energy, we place constraints on the remnant mass and
yields, and we conclude by comparing these constrained results to
current remnant mass and remnant yield observations.
 
\section{Remnant Masses}
\label{sec:remnant}

The supernova explosion directly determines the fate of the compact
remnant.  In one extreme, if an explosion is unable to launch,
material accretes onto the proto-neutron star, ultimately causing it
to collapse to form a black hole.  Without an explosion or further
ejection, the entire star accretes onto the black hole, forming a
black hole mass equal to the mass of the star at collapse.  Instead,
if a strong explosion is produced, the material above the
proto-neutron star is ejected leaving behind a neutron star with a
mass set to that of the proto-neutron star.  In between these two
extremes in the explosion energy, a range of remnant masses can be
produced.

Initial remnant mass distributions were based solely on the structure
of the progenitor star~\citep{timmes96} guided by nucleosynthetic
yield requirements.  Such estimates did not include an understanding
of the explosion mechanism itself.  More recent remnant-mass estimates
have included constraints based on the supernova engine.  For most of
these, the method has focused on increasing the energy deposition due
to neutrinos~\citep{frohlich06,fischer10,ugliano12,ertl16,sukhbold16}.
For example, \cite{perego15} tap energy from the $\mu$ and $\tau$
neutrinos, calibrating their model by fitting SN 1987A.  With these
calibrated engines, they can produce a ``best-set'' of explosion
models versus progenitor mass.  These models typically still deposit
the energy at the neutrino gain region.  In the supernova
engine, convection can redistribute the energy across the entire
convective engine.  Here we want to probe the broader range of 
explosion possibilities where we include this redistribution of the 
energy.

Initial models considering these effects were done analytically.  The
energy of the convective engine model can be estimated by determining
the energy stored in the convection region when pressure of this
region overcomes the pressure of the infalling star.  The
corresponding explosion energies from this analysis are a few times
10$^{51}$\,erg, explaining why most supernova energies also lie in
this range (even though 10$^{53}$\,erg of gravitational potential
energy is released in the collapse).  With estimates based on this
convective engine, \cite{fryer01} and \cite{fryer12} predicted the
remnant mass as a function of progenitor mass for a range of different
stellar evolution models.  Another approach has been to induce
explosions in 1-dimensional models and calculate the yields and
remnant masses from these explosions.  The problem with this latter
technique is that the results vary wildly upon how the explosion is
induced.  In this work, we strive to better understand how much the
results can vary within the convective paradigm and our parameterized
models are designed to produce a full range of results.  From this
range, we can use observational constraints to limit our explosion
models.

The processes that set the remnant mass can be separated into 3 basic
phases: the core mass post-bounce, the timing of the explosion and the
fallback post-explosion.  As electron capture produces a runaway
collapse, the core collapses.  When the core approaches nuclear
densities, strong nuclear forces and neutron degeneracy pressure halt
the infall, producing a bounce.  When the bounce shock moves outward,
it leaves behind a dense core, the seed of the proto-neutron star.
The bounce of the core depends upon the entropy and for rapidly-spinning
stars, the rotation in the core.  The masses of this initial
proto-neutron star are typically $\sim 0.9\pm0.2\,M_\odot$\citep{fryer12}.

After the stall of the bounce shock, the region between the dense core
and the stalled shock is typically unstable to Rayleigh-Taylor
instabilities.  Other instabilities can also exist, most notably the
standing accretion shock instability~\citep{houck92,blondin03}.
During this phase, material from the collapsing star is transported
down to the proto-neutron star and is assimilated onto this core.
This phase continues until either the energy in this convective region
is sufficient to drive an explosion or the proto-neutron star
collapses to form a black hole.  The longer it takes to explode, the
larger the proto-neutron star core becomes.  By modifying the onset of
the energy deposition and its power, our simulations produce a range
of remnant masses at the launch of the explosion.  If the mass exceeds
the maximum neutron star mass, the core will collapse to form a black
hole with no explosion.  The most massive stellar-mass black holes
are formed in these failed explosions.

If the energy in the convective region is sufficent to drive an
explosion, the infalling material is pushed outward.  This explosion
shock decelerates as it pushes outward, causing its velocity to drop
below the escape velocity.  This material will ultimately fall back
onto the proto-neutron star~\citep{colgate71,fryer06}.  This fallback
accretes onto the neutron star and is responsible for making high-mass
neutron stars and low-mass black holes.  We follow our explosions out
to 4000\,s at which time we can determine which material will fall
back (based on the escape velocity) and accurately calculate the final
remnant mass.

\section{Simulations}
\label{sec:simulations}

\begin{figure}[!hbtp]
\centering
\plotone{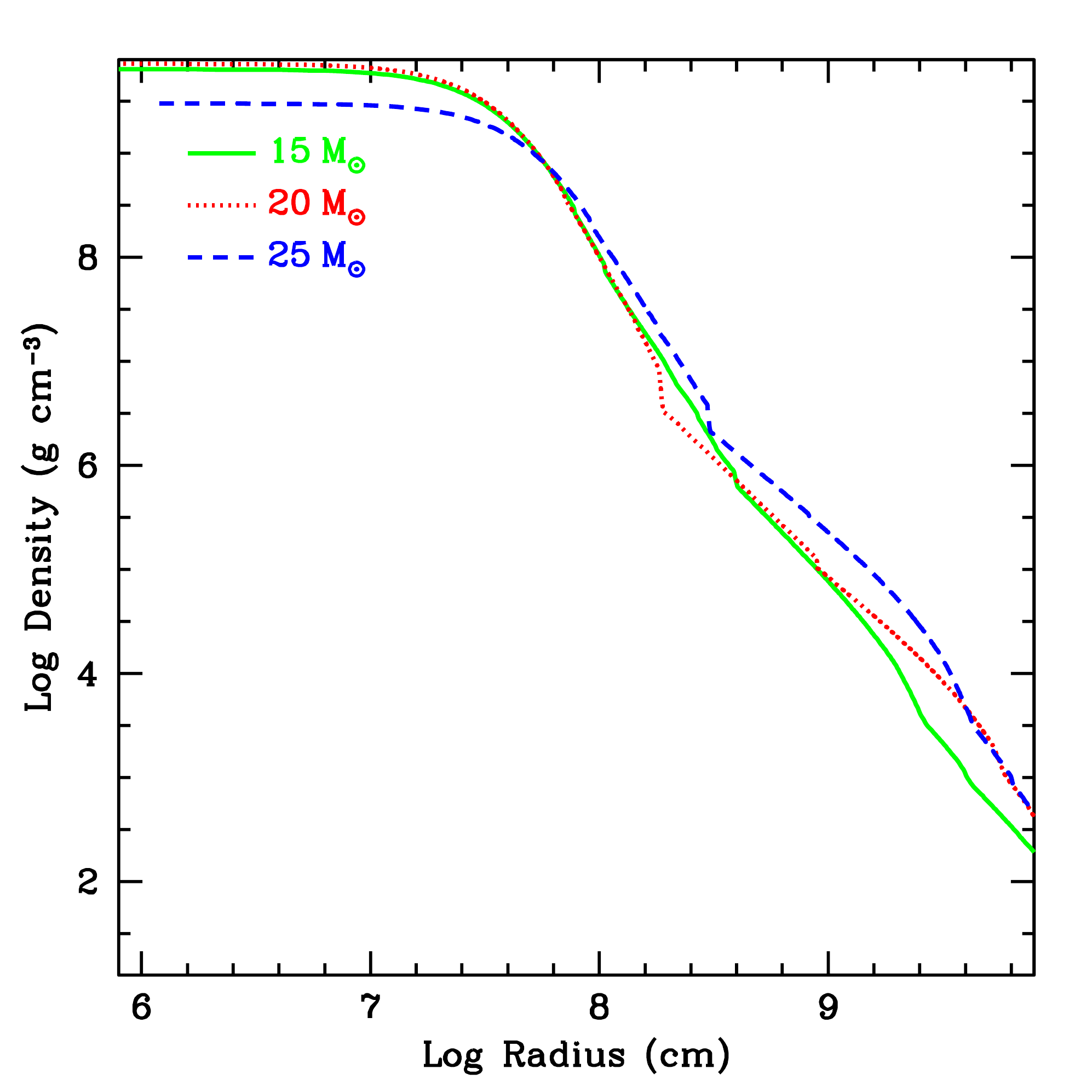}
\caption{Log density versus log radius for our 3 progenitor models:
  15\,M$_\odot$ (solid), 20\,M$_\odot$ (dotted), 25\,M$_\odot$
  (dashed).  The peak density is more a reflection of the extent of
  the collapse and less on the structure of the star. The slight
  changes in the slope between the different star shows the more
  compact structure of the 25\,M$_\odot$ star.  The decrease in the
  20\,M$_\odot$ density above $10^8$\,cm reflects a jump in the
  entropy of the star at that point.  In that sense, the 20\,M$_\odot$
  star might be easier to explode than the 15\,M$_\odot$ star.}
\label{fig:starden}
\end{figure}

In this paper, we focus on the fate of stellar collapse as a function
of the energy injection in the supernova engine.  By studying a wide
range of injection engines, we can study trends in the remnant mass
and final supernova yields as a function of explosion energy.  For
this study, we use 3 different progenitor masses (15, 20, and
25\,$M_\odot$) computed using the KEPLER
code~\citep{weaver78,woosley02,heger10}.  We chose these as
representative models for exploding systems where the convective
engine can produce a range of results.  For low-mass core-collapse
systems (roughly 8-10\,M$_\odot$), the envelope is so weakly
bound~\citep{ibeling13,woosley15} that an explosion typically occurs
quickly with little fallback~\citep{kitaura06}.  It is unlikely that
there is a lot of variation in the engine for these progenitors.
Likewise, for stars with minimal mass loss with masses above
30\,M$_\odot$, most simulations predict that the convective engine
fails to drive an
explosion~\citep{fryer99b,heger03,oconnor13,sukhbold16}, although
see~\cite{muller16}

For these calculations, a $19$-species nuclear network
\citep{weaver78} is used at low temperatures (until the end of oxygen
burning); for silicon burning band beyond, KEPLER switches to a
quasi-equilibrium approach that provides an efficient and accurate
means to treat silicon burning including convection and then
transitions to a nuclear statistical equilibrium network after silicon
depletion.  For convection we use the Ledoux criterion and mixing
length theory.  For semiconvection we use a diffusion coefficient
which is 10\% that of thermal diffusion, which roughly corresponds
to the to the efficiency resulting from the formulation by
\citet{langer83} with an $\alpha$ value of $0.04$.  We also include
thermohaline mixing according \citep{heger05}, but the process has no
major impact on the models.  All processes are formulated for use
with a general equation of state (see Appedix in \citealt{heger05})
and mixing is implemented as a diffusive processes.  For consistency
with prior work the initial composition is taken from
\citet{grevesse93}.  The models are the same as presented in
\citet{jones15} but evolved to the presupernova stage.

\begin{figure}[!hbtp]
\centering
\plotone{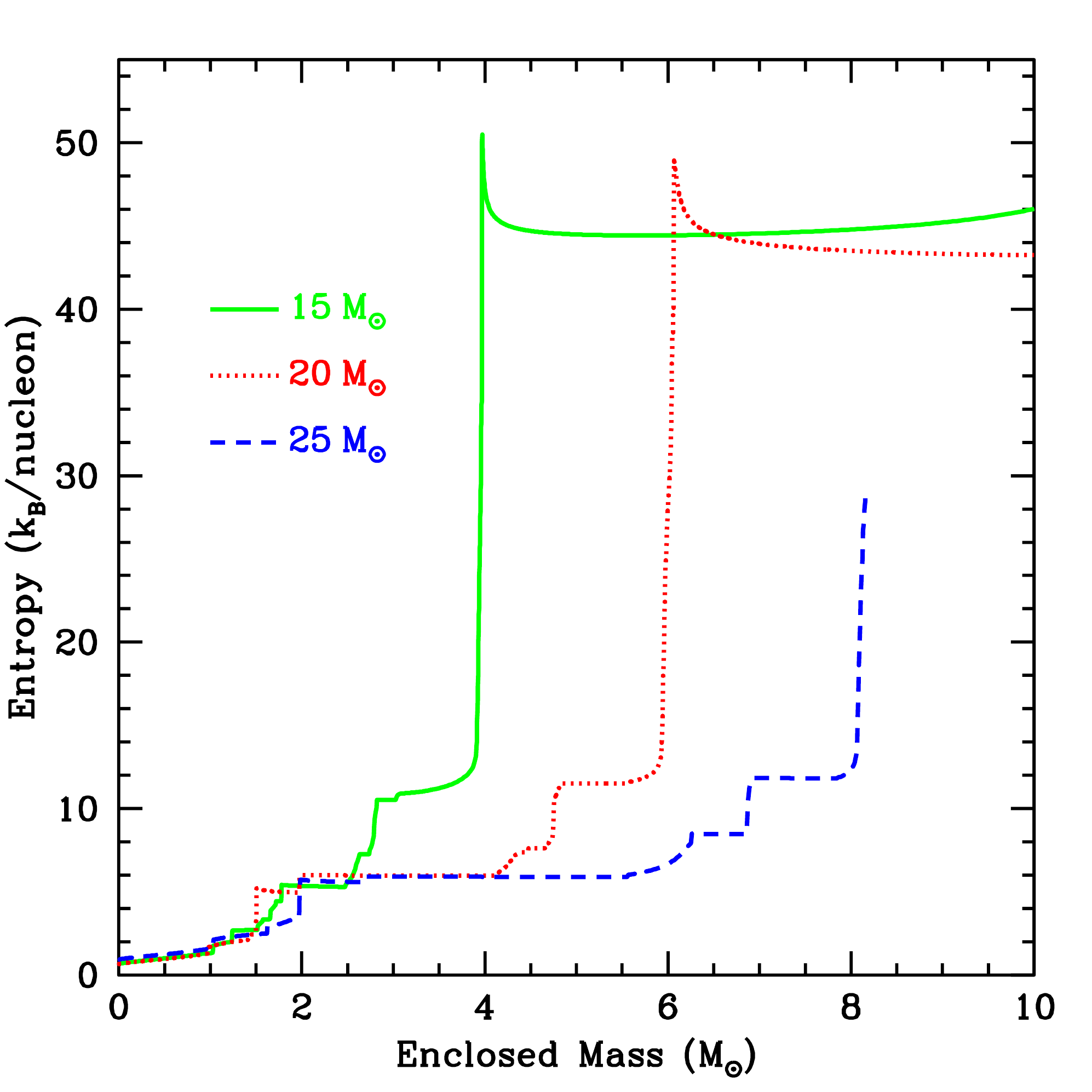}
\caption{Entropy versus enclosed mass for our 3 progenitors:
  15\,M$_\odot$ (solid), 20\,M$_\odot$ (dotted), 25\,M$_\odot$
  (dashed).  The jumps in entropy typically mark convective zone, and
  hence, composition boundaries.}
\label{fig:starent}
\end{figure}

Figure~\ref{fig:starden} shows the density profiles of our 3
progenitor stars.  Although the inner density mostly shows the extent
of the collapse (the core of the 25\,M$_\odot$ has not collapsed as
deeply into its potential well), the outer densities show the more
compact structure of the 25\,M$_\odot$ star.  The sharp decrease in
the density of the 20\,M$_\odot$ above $10^8$\,cm reflects the entropy 
jump in the star at this radius.  This is easier to see in the entropy 
profiles shown in Figure~\ref{fig:starent}.  At roughly 1.5\,M$_\odot$, the 
entropy in the 20\,M$_\odot$ rises sharply.  This corresponds to the 
lower-edge of the oxygen-burning shell which may produce the seeds 
to convection within the supernova engine~\citep{muller16}.

Our simulations are set up to mimic the basic convective-engine
paradigm with our 1-dimensional models.  In fast rotating models,
centrifugal support can slow the collapse before the material reaches
nuclear densities, causing it to bounce at a lower
density~\cite{monchmeyer89,fryer00} and our 1-dimensional models would
not capture the collapse/bounce phase.  At still higher rotation
speeds, disks can form around the collapsed core.  In both cases, the
angular momentum profiles from these collapses produce sub millisecond
pulsars and it is likely that such explosions are rare.  Unless the
core is rapidly rotating, the collapse and bounce phase can be modeled
reasonably accurately in 1-dimension.  Our 1-dimensional collapse code
utilizes a Lagrangian hydrodynamics scheme coupled to gray neutrino
transport with 3 neutrino species: electron, anti-electron, $\mu +
\tau$ neutrinos~\citep{herant94,fryer99}.  This code includes general
relativistic effects (spherically symmetric), an equation of state for
dense nuclear matter combining the Swesty-Lattimer equation of state
at high densities~\citep{lattimer91} and the Blinnikov equation of
state at low densities~\citep{blinnikov96}, and an 18-isotope nuclear
network~\citep{fryer99}.  With this code, we follow the collapse and
bounce of the core of our core-collapse progenitors.

Within the convection-enhanced neutrino-driven supernova paradigm,
energy from the hot proto-neutron star and the continually accreting
material drives convection.  If the convection is rapid, this energy
is redistributed across the entire convective region.  To mimic this
explosion process, we have introduced a series of parameters including
the region into which the energy is deposited (representing the size
of the convective region) as well as the energy deposition rate.  In
the simplest models, once the explosion is launched, energy deposition
halts.  Material can continue to accrete even after the launch of the
shock.  This fallback material can drive further outflows, depositing
additional energy~\citep{fryer09} and are seen in many
multi-dimensional core-collapse calculations~\citep[e.g.,][]{lentz15}.
To include this effect, we allow a broad range of durations, including
calculations that allow a time-dependent energy-deposition rate.
Although this allows us to include more properties of the convective
engine than past studies, bear in mind that these are still
1-dimensional simulations and do not include the full effects of
multi-dimensional calculations.

Finally, we have included a small number of models for our
most-massive progenitor that have extremely late-time energy
depositions.  Late-time energy deposition can either be through
late-time fallback or some magnetically-driven engine, e.g. magnetar.

We vary many of the parameters independently.  But, in
  an actual engine, some parameters depend upon each other.  For
  example, a higher-power drive will produce more vigorous convection
  and can produce a larger convective region.  Similarly, a weak drive
  likely ultimately produces a smaller convective region.  But other
  factors (like the initial seeds and rotation) can also affect the
  size of the convective region.  Hence, for this paper, we vary these
  parameters independently.  Likewise, a strong explosion is likely to
  have a shorter deposition time.  Our parameter space allows a strong
  explosion with a long deposition time.  In this manner, we produce
  stronger explosions than one would expect with the convective
  engine.  In such cases, our models do not represent the classic
  convective engine alone, rather mimicking magentically-driven or fallback
  energy sources.  If we limited the results to the classic convective
  engine, we could constrain our parameters somewhat.

\section{Results}
\label{sec:results}

\subsection{Compact Remnant Masses}

\begin{figure}[!hbtp]
\centering
\plotone{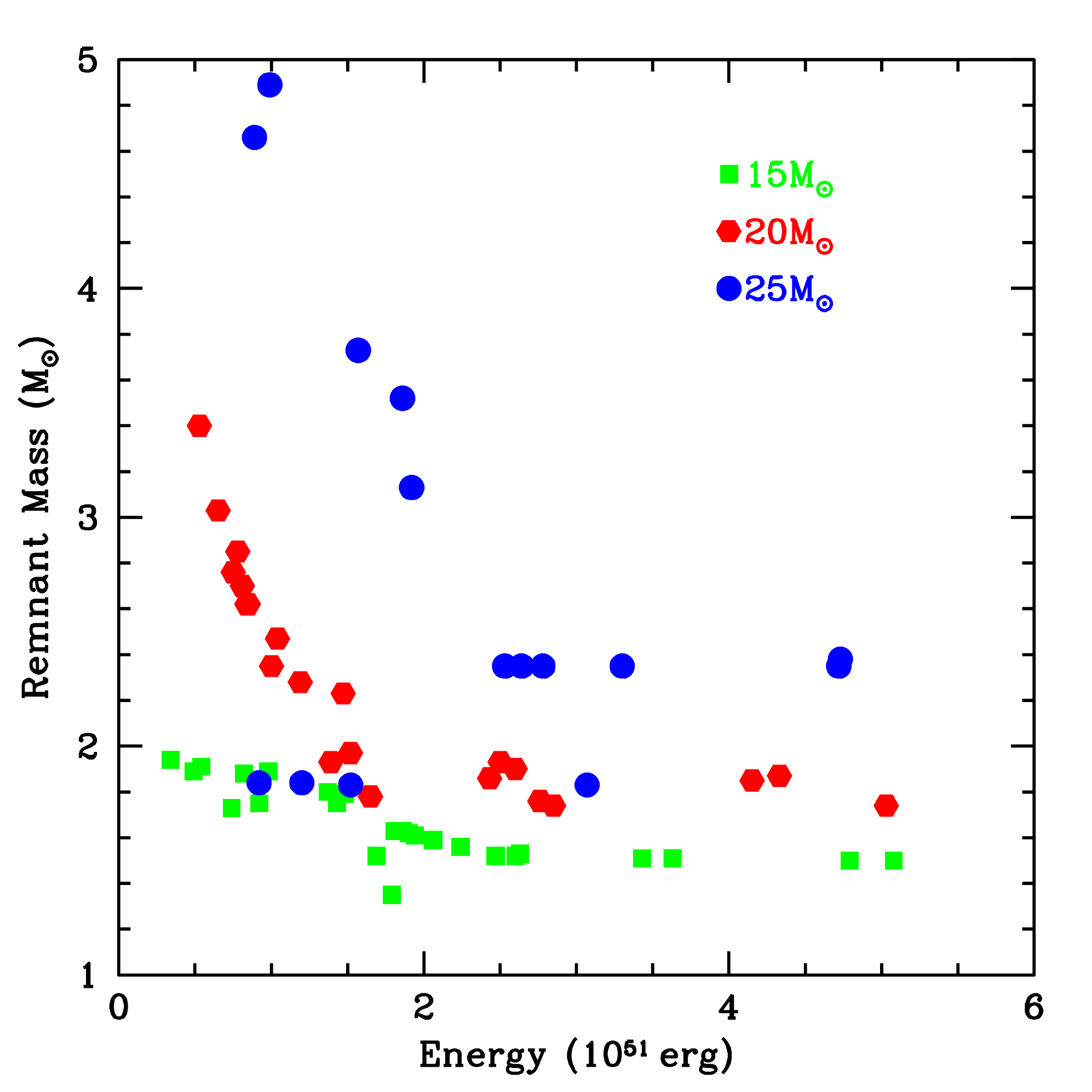}
\caption{Remnant mass versus explosion energy for our 3 progenitor
  masses: 15\,M$_\odot$ (square), 20\,M$_\odot$ (hexagon),
  25\,M$_\odot$ (circle).  }
\label{fig:remnant}
\end{figure}

Ideally, for a given progenitor star, there would be a one-to-one
correspondence between explosion energy and remnant mass.
The convective engine depends upon the growth of the
  convective instabilities that, in turn, depend upon a variety of
  features of the progenitor, including the progenitor's rotation and
  the magnitude of the turbulence in the progenitor's burning layers.
  The asphericities in the burning layers provide the seeds for the
  convection in the engine.  These seemingly small differences can
  cause large differences in the growth of the convective motions
  above the proto-neutron star, leading to a wide range of explosion
  energy for roughly similar progenitors: see, for example,
  \cite{couch15}.  Therefore, it is possible that progenitors of equal
  mass can produce very different explosion energies.

The explosion energy is set to the kinetic energy minus the absolute
value of the gravitational potential energy 4000\,s into the
explosion.  But because the nature of the explosion can vary wildly,
this may not be the case.  Table~\ref{tab:runs} shows the different
choices for the energy deposition and the resultant compact masses.
There are multiple injection parameters that produce roughly the same
explosion energy.  In this manner, we can determine
  whether the yields only depend upon the explosion energy and
  progenitor, or whether the nature of the explosion can alter the
  yield.

Figure~\ref{fig:remnant} plots remnant
mass versus explosion energy for these models.  These masses are given
in bayronic mass of the collapsing core.  The gravitational mass of
the remnant is likely to be $\sim$10-15\% lower.  Hence, our
15\,M$_\odot$ progenitor produces remnants with gravitational masses
in the 1.2-1.7\,M$_\odot$ range.  More massive progenitors generally
produce more massive remnants~\citep{fryer99b}: for explosions between
$1-2\times10^{51}$\,erg, remnant masses can be as high as 2\,M$_\odot$
for a 20\,M$_\odot$ progenitor and over 4\,M$_\odot$ for a
25\,M$_\odot$ progenitor\footnote{Note that if mass loss is extensive
  such that the core is modified, more massive stars can end their
  lives with small cores and hence will produce lower-mass remnants.}

Because fallback is less in stronger explosions~\citep{fryer12}, there
is a clear trend where more energetic explosions produce lower mass
remnants.  Similarly, the high binding energy of more massive stars
means that, for a fixed explosion energy, the more massive progenitors
generally produce more massive remnants.  Indeed, with our normal
explosion methods, the remnant mass for our 25\,M$_\odot$ progenitor
with ``standard'' supernova explosion energies of $1-2\times
10^{51}$\,erg produce remnants with gravitational masses ranging from
2 to 4.25\,M$_\odot$.  For a few times $10^{50}$\,erg explosion, most
of the 25\,M$_\odot$ star remains bound, making a large black hole.

Table~\ref{tab:runs} shows the wide range of total 
injection energies in our models.  As energy is injected into 
the layers above the proto-neutron star, this region begins to 
emit neutrinos copiously.  Hence, the injection energy can be 
much higher than the final explosion energy.  Explosion energies 
above a few times $10^{51}$\,erg are difficult to achieve without 
extremely high injection energies and it is unlikely that these 
high energies occur without a different engine.

Recent results from \cite{sukhbold16} argue that they can make
$<2$\,M$_\odot$ remnants with progenitor systems with initial masses
above 20\,M$_\odot$ with $10^{51}$\,erg explosions.  These explosions
are driven by enhanced neutrino energy deposition.  The neutrino
deposition region is typically very close to the surface of the
proto-neutron star and these models closely mimic our small
injection-region models.  However, our standard explosions, with
energy injections less than 1\,s were unable to reproduce these
results.  Either these progenitors are very different than ours (have
much less compact cores) or our models are missing some aspect of the
explosion.  One solution is that late-time energy injection prevents
the fallback\footnote{Note that the models of \cite{muller16b} do not
  include fallback and hence their low energy, low remnant mass
  companions are probably just an artifact of this assumption.}.  We
were able to reproduce more modest remnant masses by constructing
explosions with long-term energy injection.  In a scenario where
continued accretion through fallback or magnetar dipole radiation
inject energy at late times, we can minimize the fallback and the
final remnant mass.  In such a scenario, we can produce gravitational
masses below 1.6\,M$_\odot$ even for $<10^{51}$\,erg explosions.

Our progenitors are all modeled as single stars.  Wind mass loss
ejects half of the mass in our 25\,M$_\odot$ star, 7\,M$_\odot$ of our
20\,M$_\odot$ star and 4.5\,M$_\odot$ in our 15\,M$_\odot$ star.
Binary mass transfer can remove the hydrogen and even helium envelopes
allowing further mass loss through winds.  If this mass loss affects
the core, it will alter the fate of the collapse.  If it does not
alter the inner $\sim$3\,M$_\odot$, the explosion is not affected by
the mass loss and, unless there is considerable fallback, neither is
the remnant mass.  If mass loss does alter the core structure (e.g.
it occurs before helium depletion in the core) or fallback is
extensive, binary mass transfer can dictate the final remnant mass.

Our predictions include a few additional uncertainties.  We choose
when to start driving the energy after the bounce - the timing is
determined by the instability growth time.  This reflects the time for
the convection to develop between the proto-neutron star and the
stalled shock.  Altering this can change the final remnant mass by
0.1-0.2\,M$_\odot$.  In addition, after the launch of the shock, we
``accrete'' onto the proto-neutron star after the material exceeds a
density of between 10$^{10}-10^{12}$\,g\,cm$^{-3}$, arguing that above
this density, neutrino cooling is rapid, allowing it to accrete
quickly.  For models with over a solar mass of fallback and late-time
drives, this density limit can also change the mass by
0.1-0.2\,M$_\odot$ (the mass increases with a larger density limit).

\subsection{Basic Yields}

The yields from a given progenitor are also sensitive to the details
of the explosion.  For this paper, we use the publicly available TORCH
code~\citep{timmes00} ({\it http://cococubed.asu.edu}) to post-process
our ejecta trajectories and calculate a few basic yields from the
supernova explosion: oxygen, neon, magnesium, silicon, sulfur, argon,
calcium, and iron peak elements using a 489 isotope network.  The
initial abundances are taken from the network in our stellar models,
limited to 19 isotopes (mostly alpha-chain, including the 8 isotopes
considered here).  For these nuclear network calculations,
we use the first 2000 (equally-spaced, 10\,ms) timesteps, following
the ejecta for 20\,s, well after nuclear burning is complete.  
The Torch code sub-samples these time dumps if more resolution is
needed.  Both the choice of the initial abundances and this time
sampling will be discussed in more detail in a later paper (Andrews et
al., in preparation).  These calculations are intended to demonstrate
the range in yields.  The exact yields will vary with a more complete
isotope distribution in the initial conditions and understanding the
nucleosynthetic yield errors must also include errors in the initial
composition, time sampling, network conditions (see Andrews et al., in
preparation).

\begin{figure}[!hbtp]
\centering
\plotone{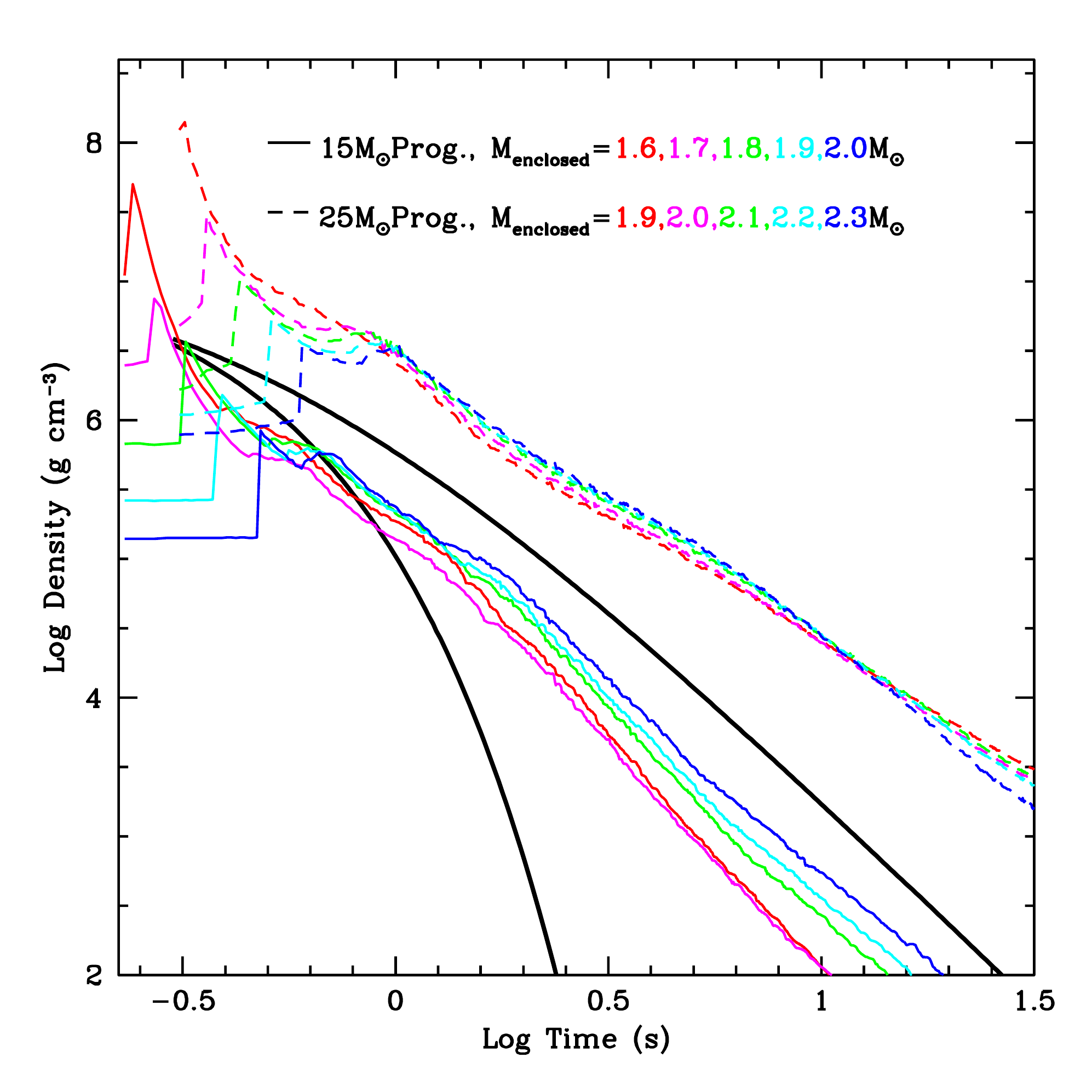}
\plotone{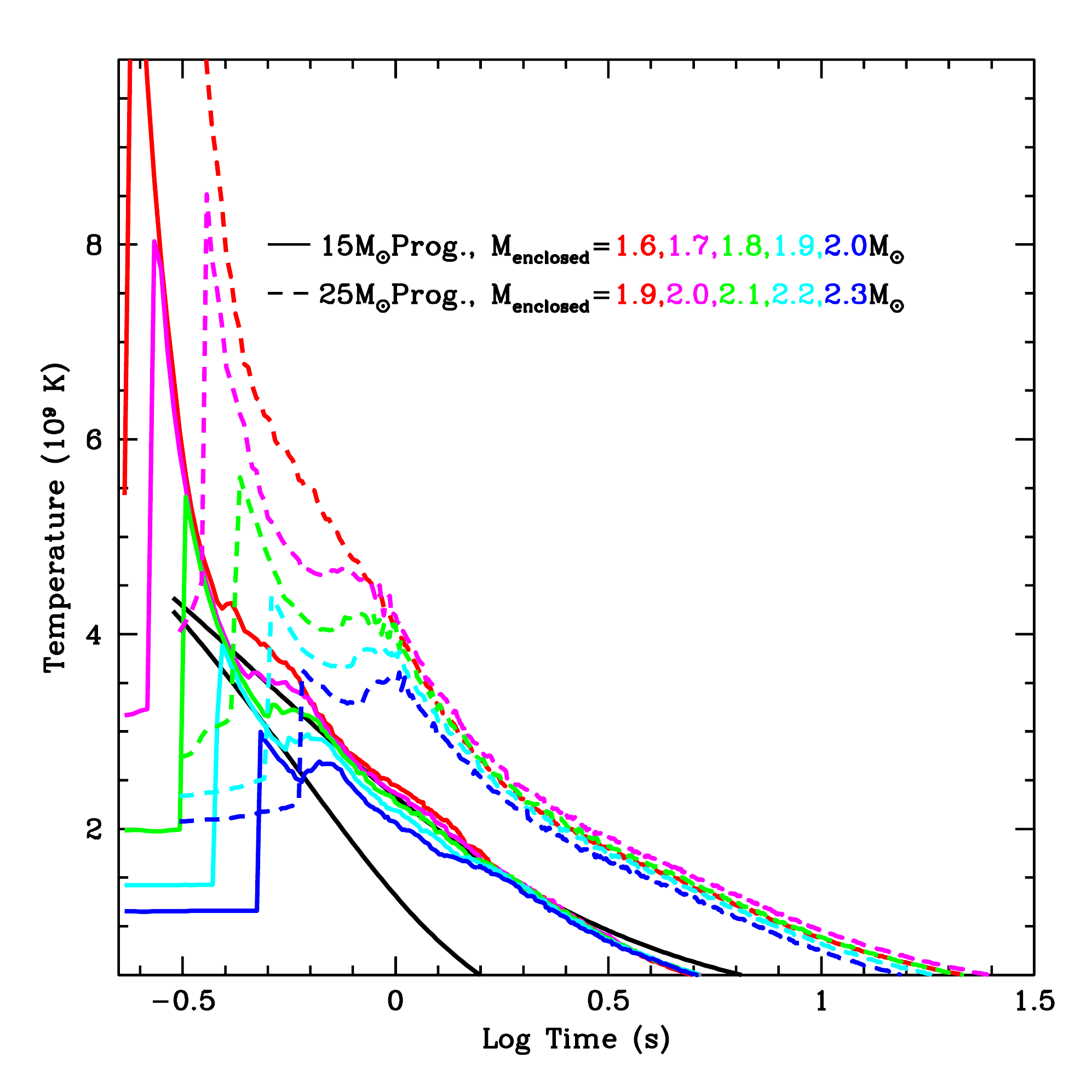}
\caption{Density and temperature versus time for
  ejected matter.  These trajectories are taken for matter at a range
  of positions (denoted by the enclosed mass) within the star with
  mass coordinates ranging from the base of the ejected material
  through the oxygen layer (the innermost layers have the highest peak
  densities and temperatures).  Note that additional shocks can both
  increase the density and temperature with time and the trajectories
  are not necessarily monotonic.  On top of these curves are the
  standard analytic fits to the trajectory evolution.}
\label{fig:exp}
\end{figure}

As the supernova explosion plows through the star, the shock hitting
the stellar material accelerates it while compressing and heating it.
The shock-heated material then adiabatically cools as the star
expands.  The final yields are set by the peak temperature and density 
at this peak temperature as well as the timescale of the cooling.  Typically, 
astronomers use one of two profiles, both assuming adiabatic expansion, but 
with different expansion evolutions\citep{magkotsios10,harris17}.  Originally, 
an exponential decay was assumed\citep{hoyle64,fowler64}:
\begin{equation}
T=T_0 e^{-t/\tau}
\end{equation}
and
\begin{equation}
\rho=\rho_0 e^{-t/3\tau}
\end{equation}
where the pre-shock temperature and density are $T_0$ and $\rho_0$ 
and a decay time $\tau = (446/\rho_0^{0.5})$.  Comparisons to 
explosion calculations suggests an alternative, constant-velocity 
expansion that produces a power-law evolution\citep{magkotsios10} 
similar to some wind calculations\citep{panov09}:
\begin{equation}
T=T_0/(2 t + 1)
\end{equation}
and
\begin{equation}
\rho=\rho_0/(2 t + 1)^3.
\end{equation}
Figure~\ref{fig:exp} shows temperature and density profiles for a 
set of mass points for two different explosions (strong explosions 
of a 15\,M$_\odot$ and 25\,M$_\odot$ progenitor).  Although the 
power law evolution is the better fit to the data, especially after 
the ejecta drops out of nuclear statistical equilibrium where the 
time evolution is most critical, neither analytic model captures 
all aspects of the expansion.  In particular, a succession of 
shocks can actually cause the temperature and density to increase 
as the material expands and the evolution is not necessarily monotonic.

The yields for our models are listed in Table~\ref{tab:yields}.  The
star itself produces many of the yields studied here during its
lifetime.  One of the affects of the supernova explosion is to
determine what material is ejected and what remains part of the
compact remnant.  But the shocks in the supernovae can drive nuclear
burning, further altering the abundances in the star, both
destroying and producing the elements in this study.  For example,
when the supernova shock hits the silicon layer, it drives fusion,
producing iron and destroying silicon.  Silicon, on the other hand,
can be produced when the shock heats the oxygen shell.

\begin{figure}[!hbtp]
\centering
\plotone{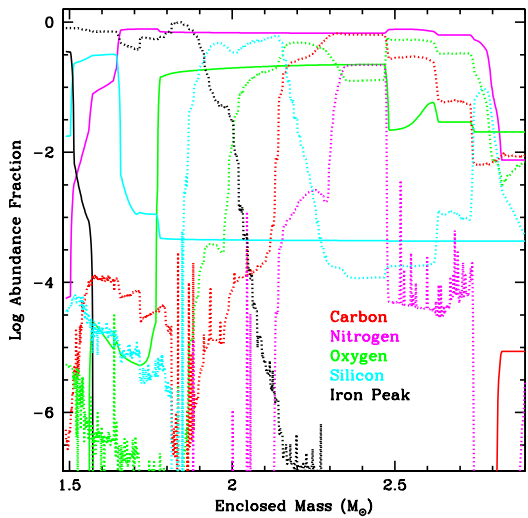}
\caption{Abundance fractions of carbon, nitrogen, oxygen, silicon and
  iron peak elements as a function of enclosed mass both at collapse
  (solid) and after the supernova explosion (dotted) for model
  M15bE5.08: a 15\,M$_\odot$ progenitor with a 5 foe
  explosion (see Table~\ref{tab:runs} for characteristics).  The strong
  shock in this explosion produces roughly 0.5M$_\odot$ of iron peak
  elements.}
\label{fig:yvmstrong}
\end{figure}

\begin{figure}[!hbtp]
\centering
\plotone{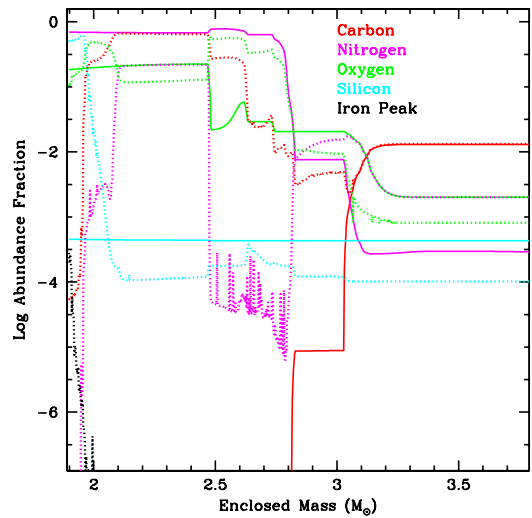}
\caption{Abundance fractions of carbon, nitrogen, oxygen, silicon and
  iron peak elements as a function of enclosed mass both at collapse
  (solid) and after the supernova explosion (dotted) for model
  M15aE0.54: a 15\,M$_\odot$ progenitor with a 0.5 foe explosion (see
  Table~\ref{tab:runs} for characteristics).  Even in this weak
  explosion where much of the inner region falls back on the compact
  remnant, the explosion resets the innermost ejecta.}
\label{fig:yvmweak}
\end{figure}

Stellar yields, shock-driven nuclear burning in the supernova, and the amount of
ejected material all play an important role in the nucleosynthetic
yields of supernovaee.  To compare the different roles these 3 factors
have on the nucleosynthetic yields, we focus on two explosions of our
25\,M$_\odot$ progenitor.  Figure~\ref{fig:yvmstrong} shows the
abundance fraction of carbon, nitrogen, oxygen, silicon and iron peak
elements for model M15bE5.08 (a 15\,M$_\odot$, $5\times10^{51}$\,erg
explosion, see Table~\ref{tab:runs} for characteristics) at collapse
and after the supernova shock has passed through it.  Although the
pre-collapse star sets the initial seeds for nuclear burning, this
strong explosion alters the abundance fraction for the entire inner
material in the star, producing roughly 0.5M$_\odot$ of iron peak
elements.  In the weak explosion (model M15aE0.54: a 15\,M$_\odot$,
$0.5\times10^{51}$\,erg explosion), much of the material synthesized
in the shock falls back onto the newly-formed neutron star
(Figure~\ref{fig:yvmweak}).  The innermost ejecta are also determined
by the supernova shock.  In both cases, the outermost material is set
by the pre-collapse progenitor.

\begin{figure*}[!hbtp]
\centering
\epsscale{0.85}
\plottwo{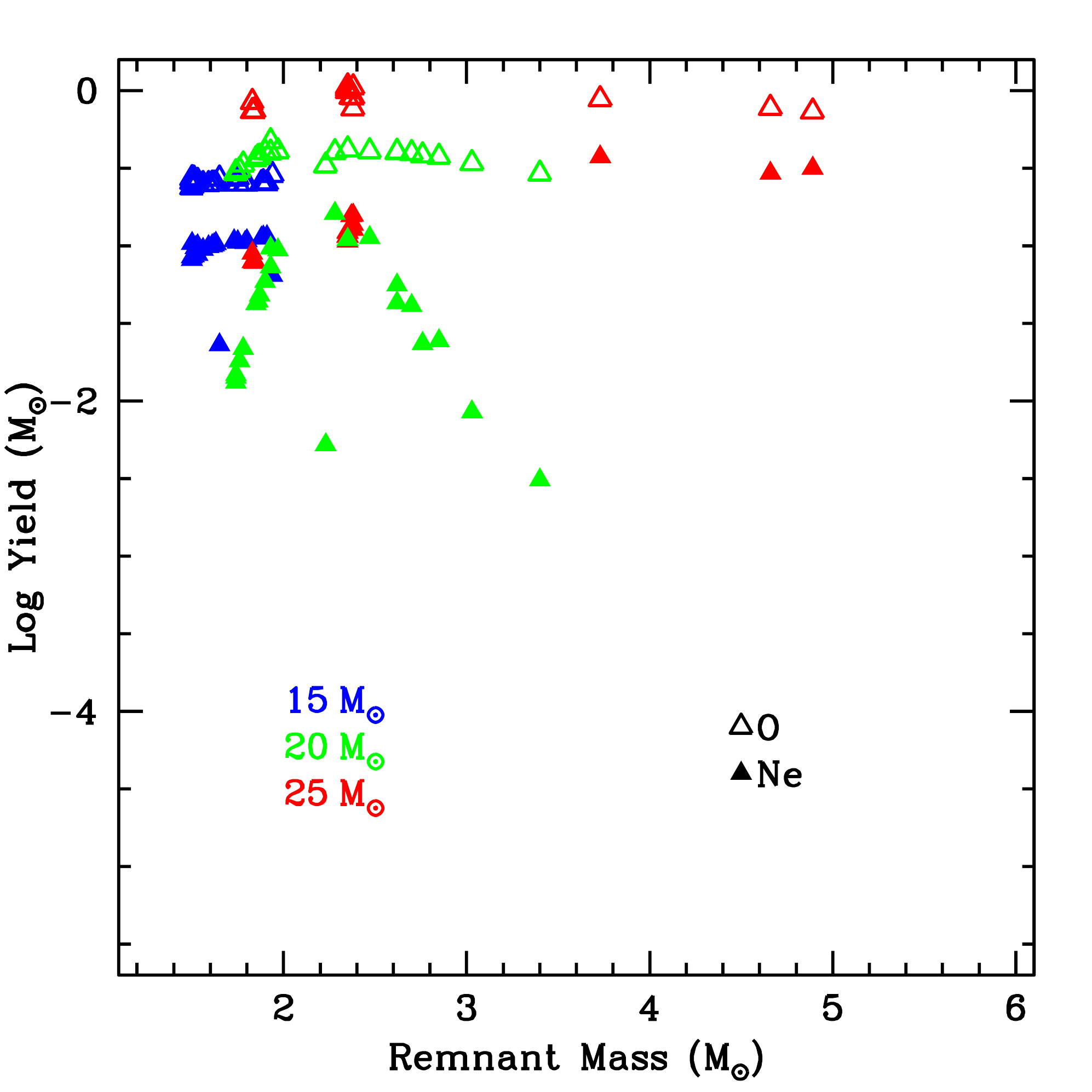}{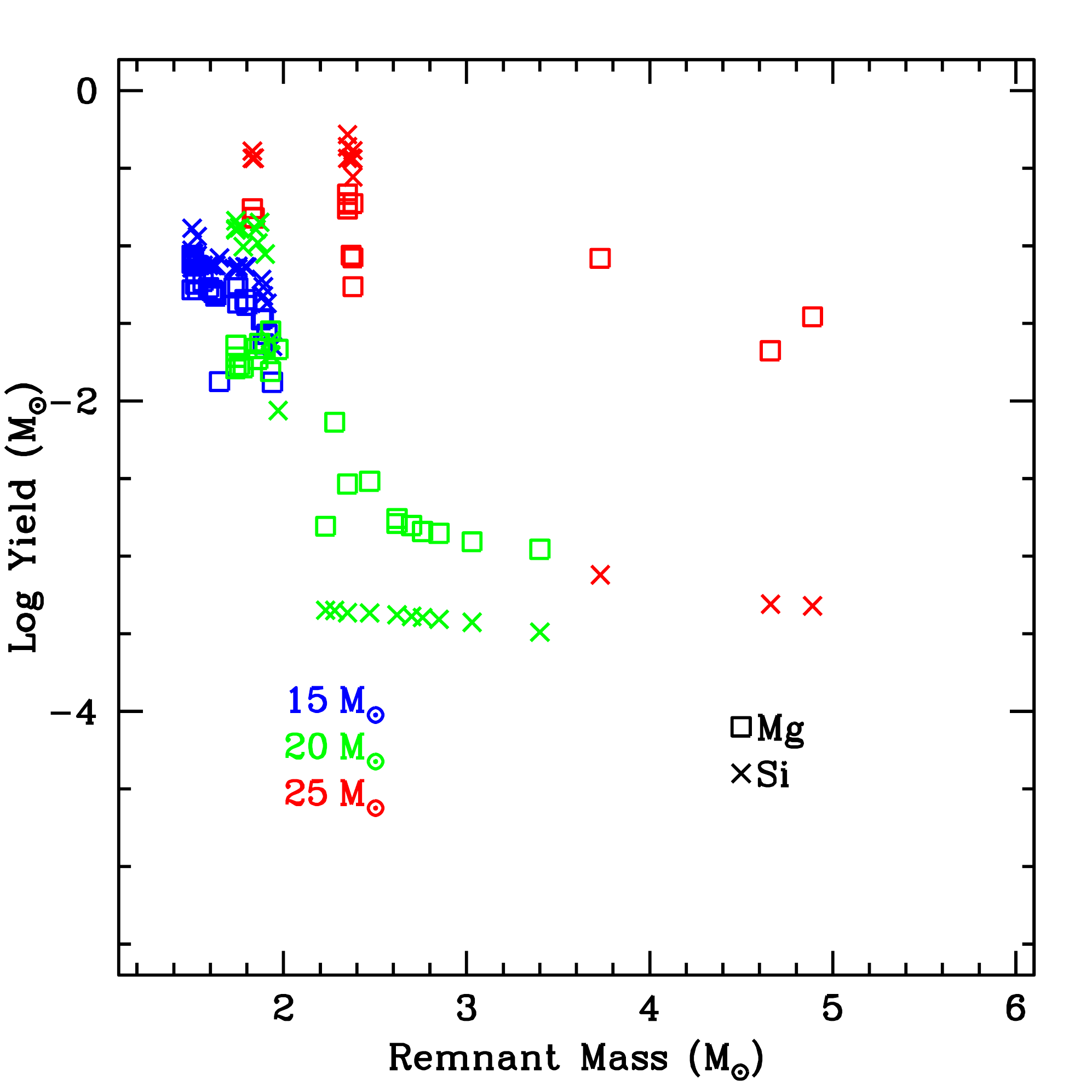}
\plottwo{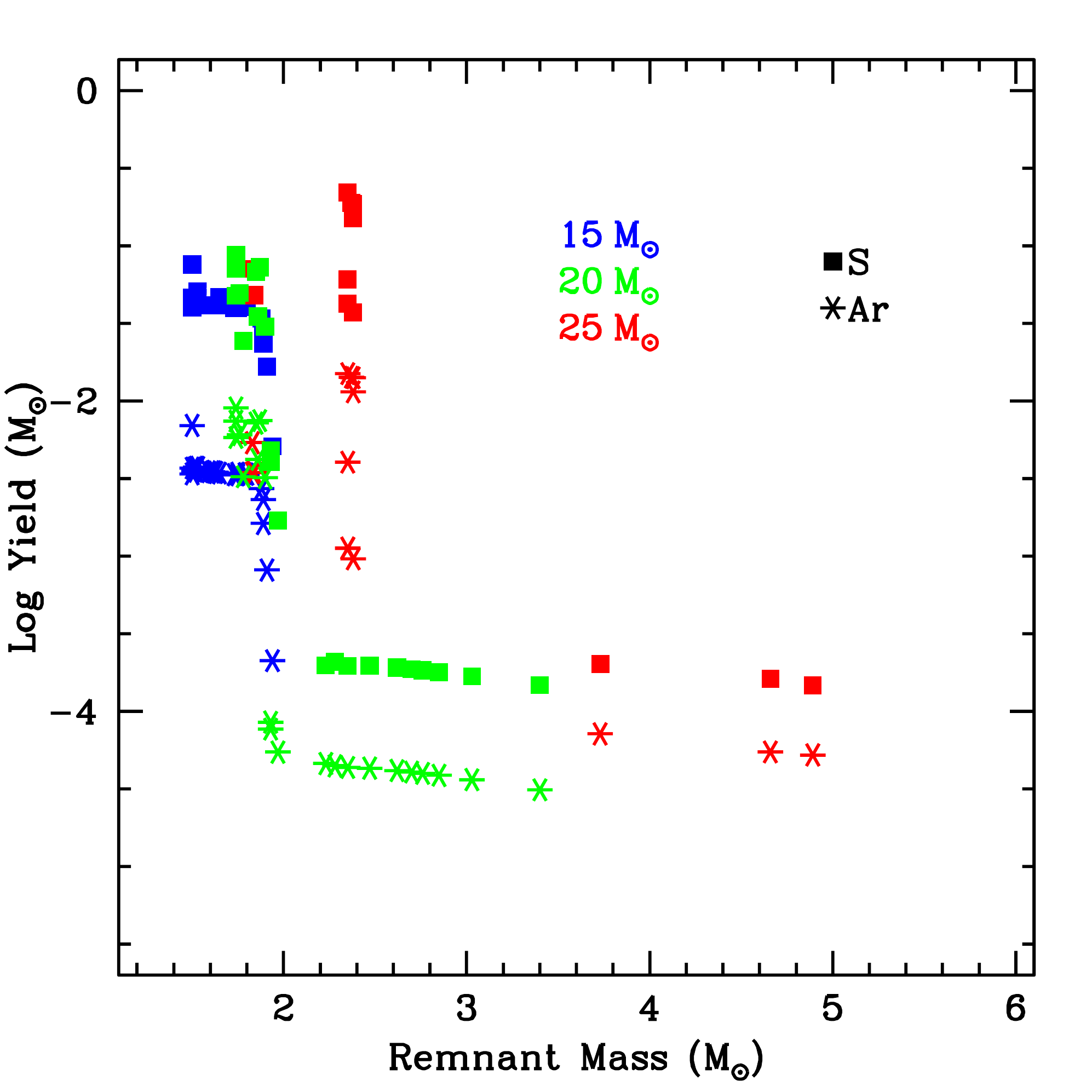}{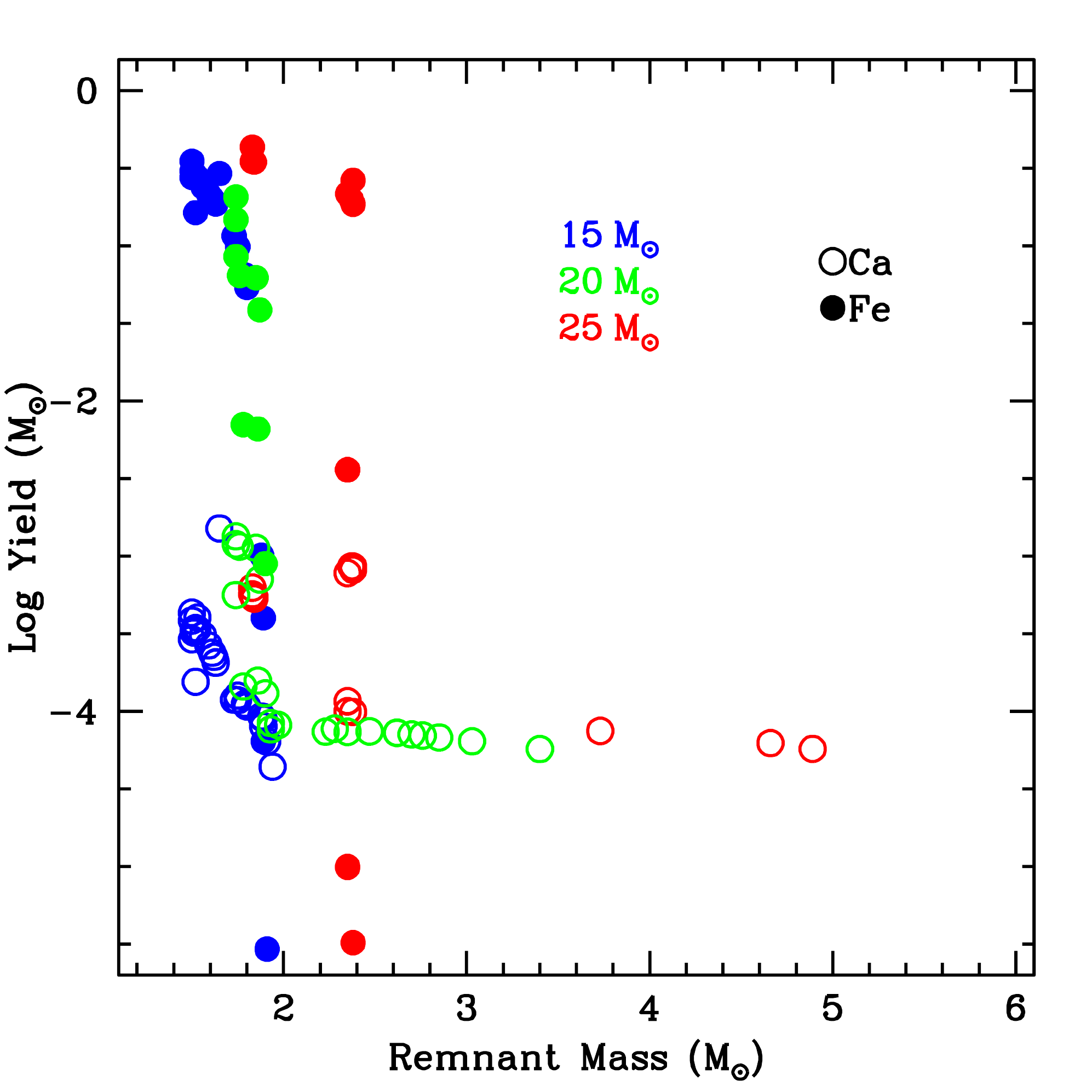}
\epsscale{1.0}
\caption{Yields versus remnant mass for our 3 progenitors:
  $15\,M_\odot$ (blue), $20\,M_\odot$ (green), $25\,M_\odot$ (red).
  The four panels show all of the yields: upper left panel shows
  oxygen (open triangle) and neon (filled triangle), lower left panel
  shows magnesium (open square) and silicon (cross), upper right panel
  shows sulfar (filled square) and argon (star), and lower right panel
  shows calcium (open circle) and iron peak (filled circle).  The
  precipitous drop in heavy element yields occurs at the top of the
  silicon shell.  This shell is shocked to form the heavy elements.
  If none of this is ejected, we eject no heavy elements.  Note,
  however, that if the explosion were asymmetric, it would be possible
  to eject some heavy elements at this remnant mass.}
\label{fig:yvm}
\end{figure*}

Figure~\ref{fig:yvm} shows the yields from explosions versus compact
remnant mass, color coded by the progenitor zero-age main sequence
mass.  Because the matter is not ejected, the yields decrease with
increasing remnant mass, especially for the heavier elements.  But
there are noticeable deviations from this trend.  First and foremost,
because the shell burning layers have more mass in the larger
progenitors, they can eject more mass, even when the remnant mass is
larger.  This is most evident in the $25\,M_\odot$ progenitor.  For
remnants less than $2.5\,M_\odot$, explosions of this progenitor can
produce more mass of all the elements in our study.  The effect of the
supernova-induced nuclear burning is also evident in the range of
yields for a given progenitor with the same final compact remnant
mass.  A strong impulse shock will produce more heavy elements than a
slower, but continuously driven shock that drives an explosion, but
prevents much fallback.

\begin{figure*}[!hbtp]
\centering
\epsscale{0.85}
\plottwo{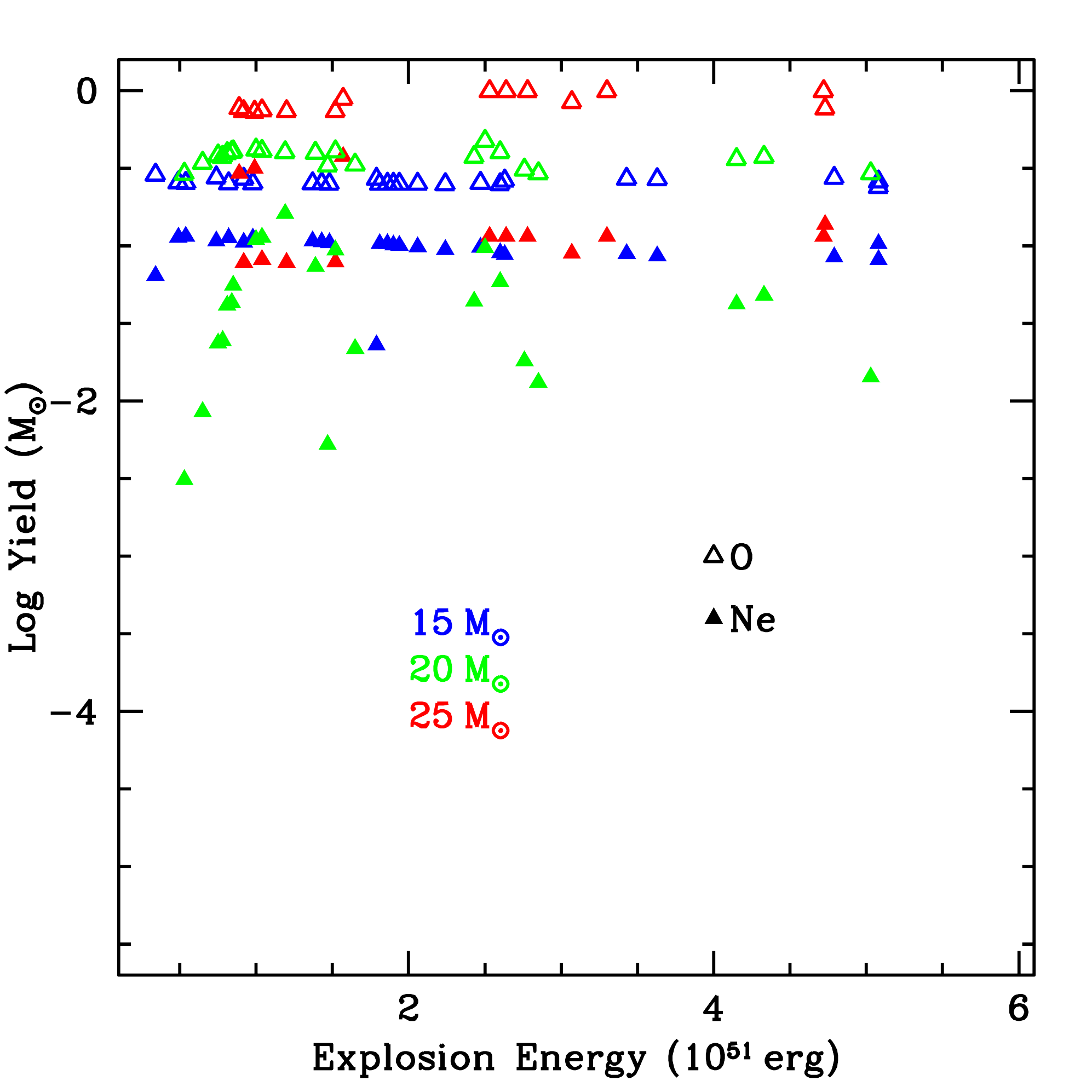}{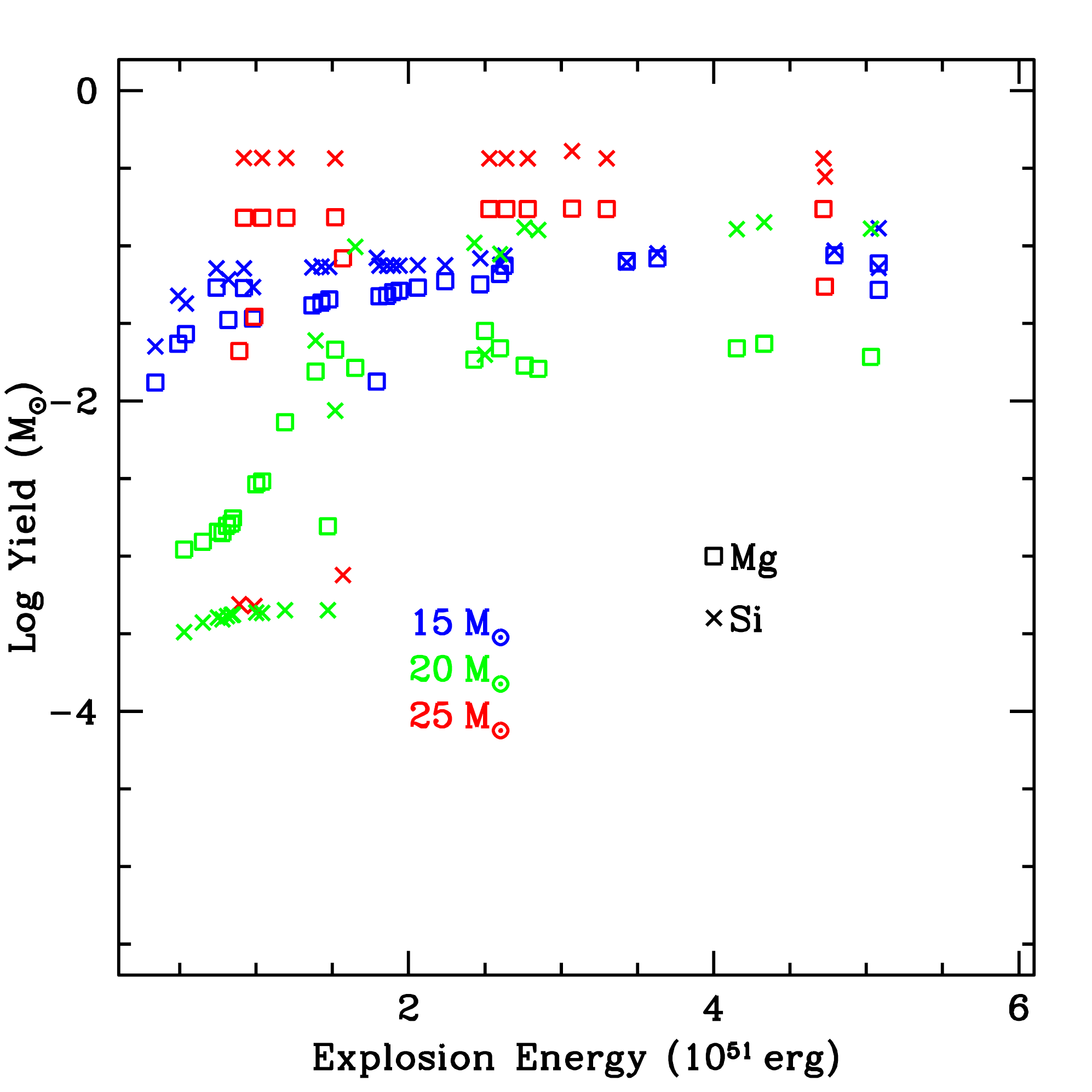}
\plottwo{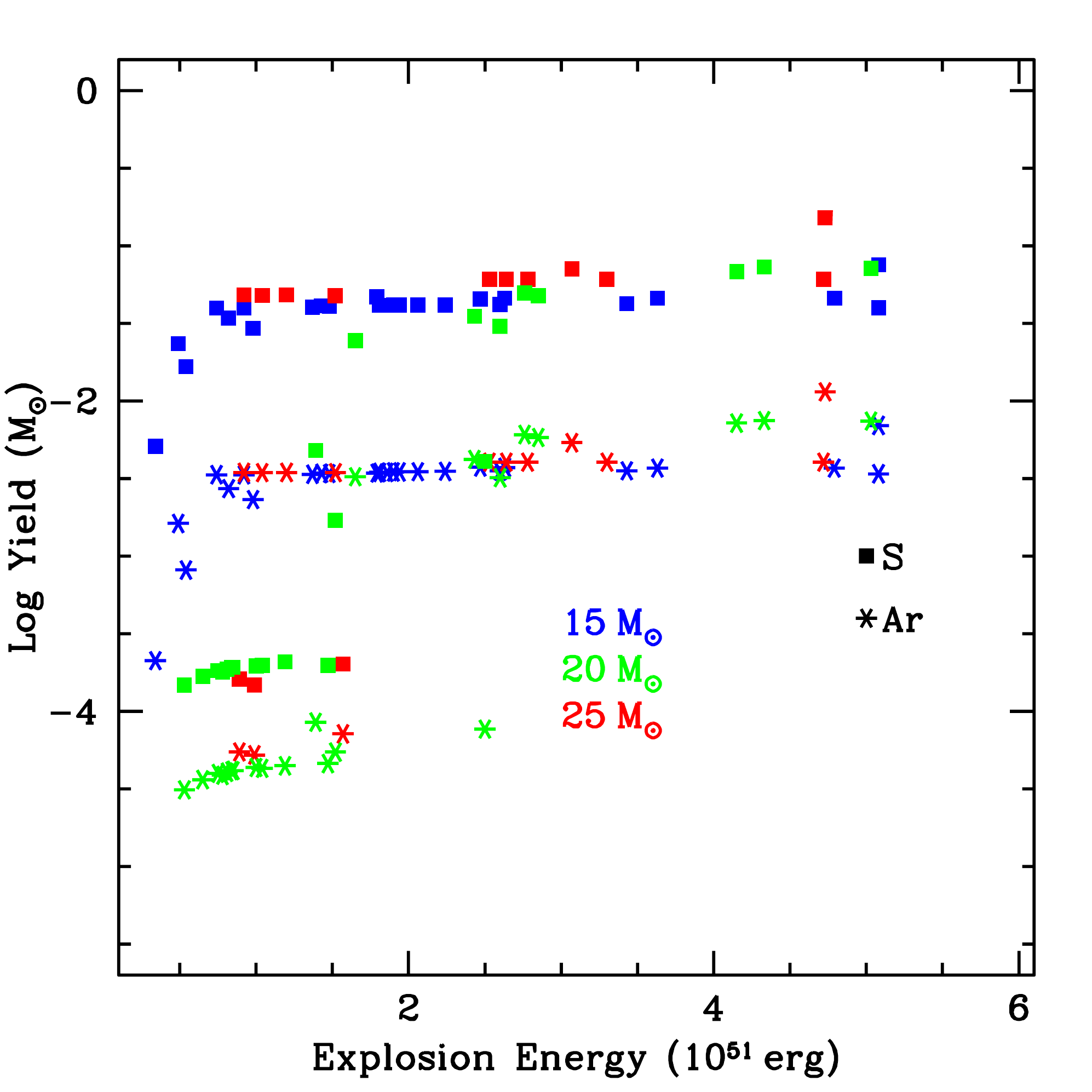}{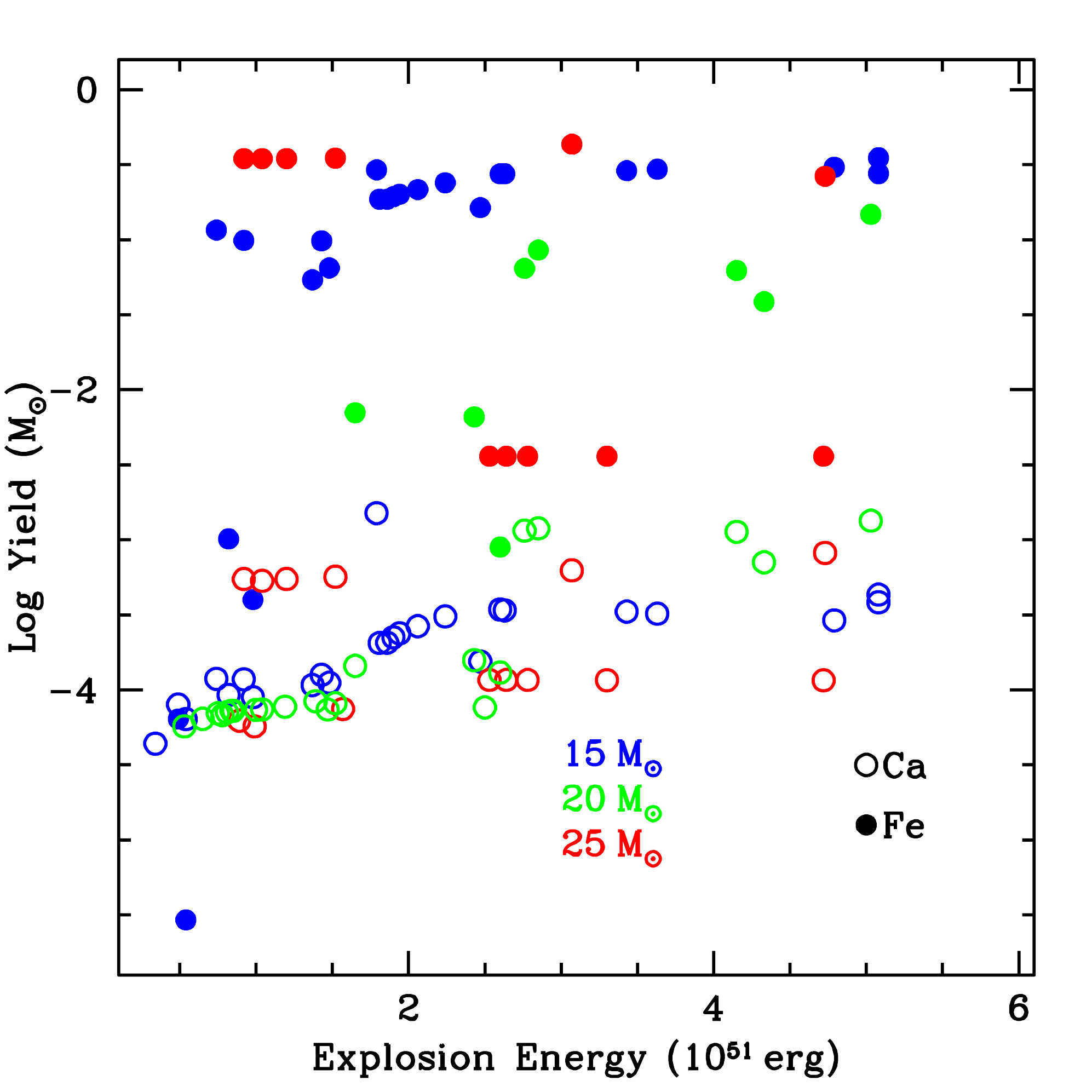}
\epsscale{1.0}
\caption{Yields versus explosion energy for our 3 progenitors with the 
same color coding and symbols as figure~\ref{fig:yvm}.} 
\label{fig:yve}
\end{figure*}

To better see the dependence of the yields on the explosion energy, we
plot the yields versus explosion energy (Fig.~\ref{fig:yve}).  At low
energies, the primary effect on the yield is the amount of fallback.
Stronger shocks (higher energy) have less fallback as most of the
material is moving well over the escape velocity.  But a continously
driven explosion can have very little fallback even with an explosion
of modest energy.  But the shock also dictates the amount of
shock-induced burning.  Especially in the $15\,M_\odot$ progenitor
where remnant masses do not change much for explosion energies above
$10^{51}$\,erg, the trends in explosion energy are more evident: neon
and oxygen are typically destroyed in stronger explosions, but many of
the other elements increase slightly.

\begin{figure*}[!hbtp]
\centering
\epsscale{0.85}
\plottwo{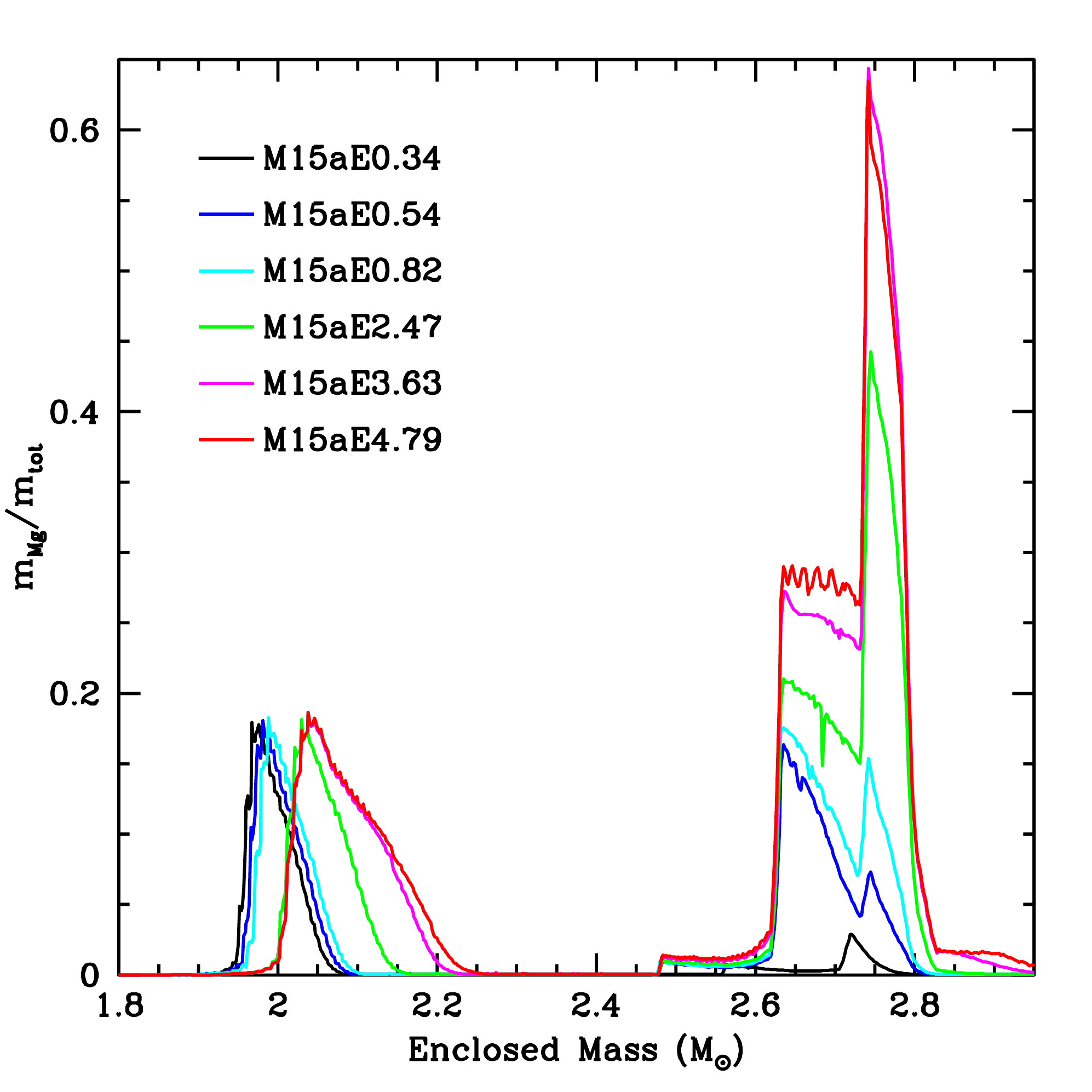}{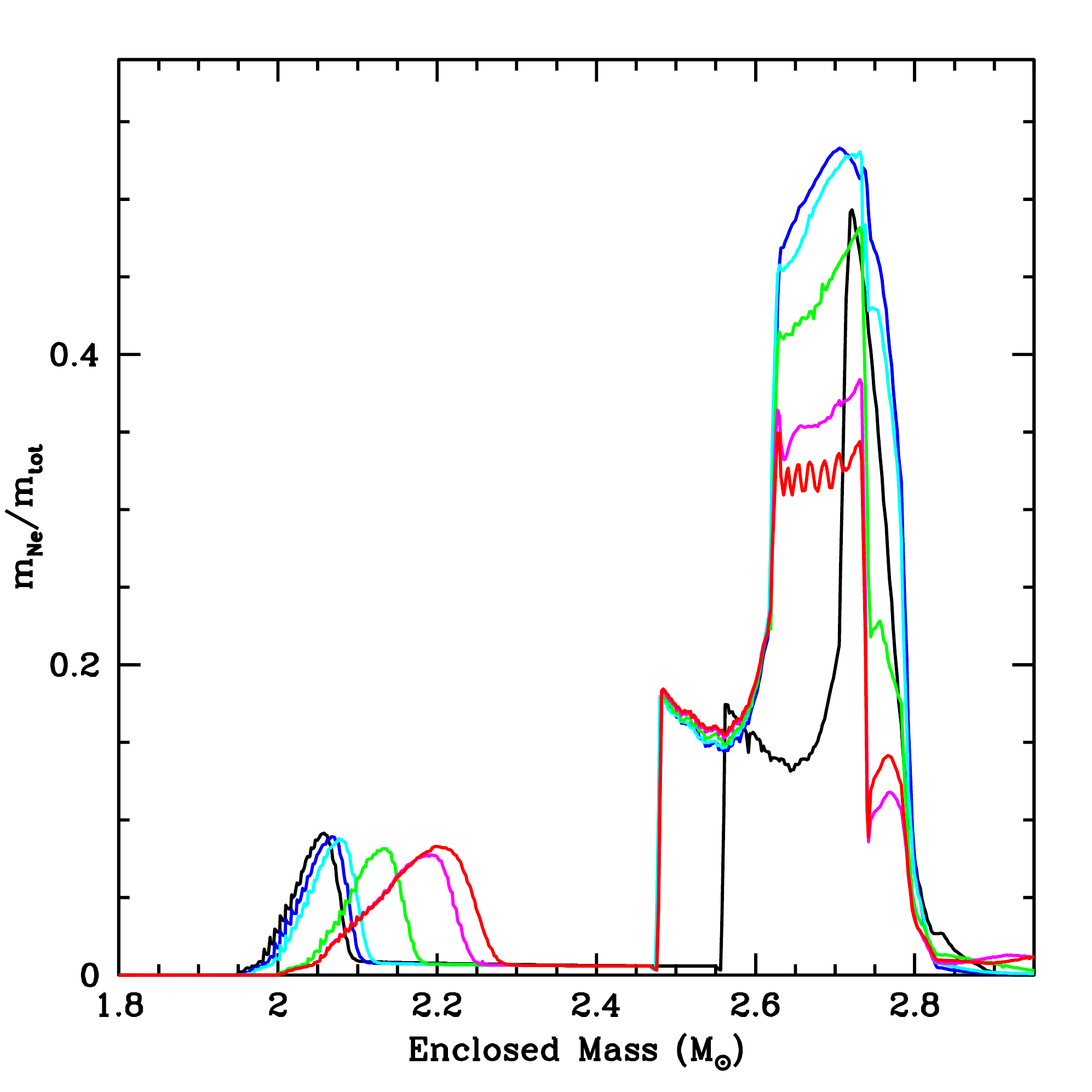}
\plottwo{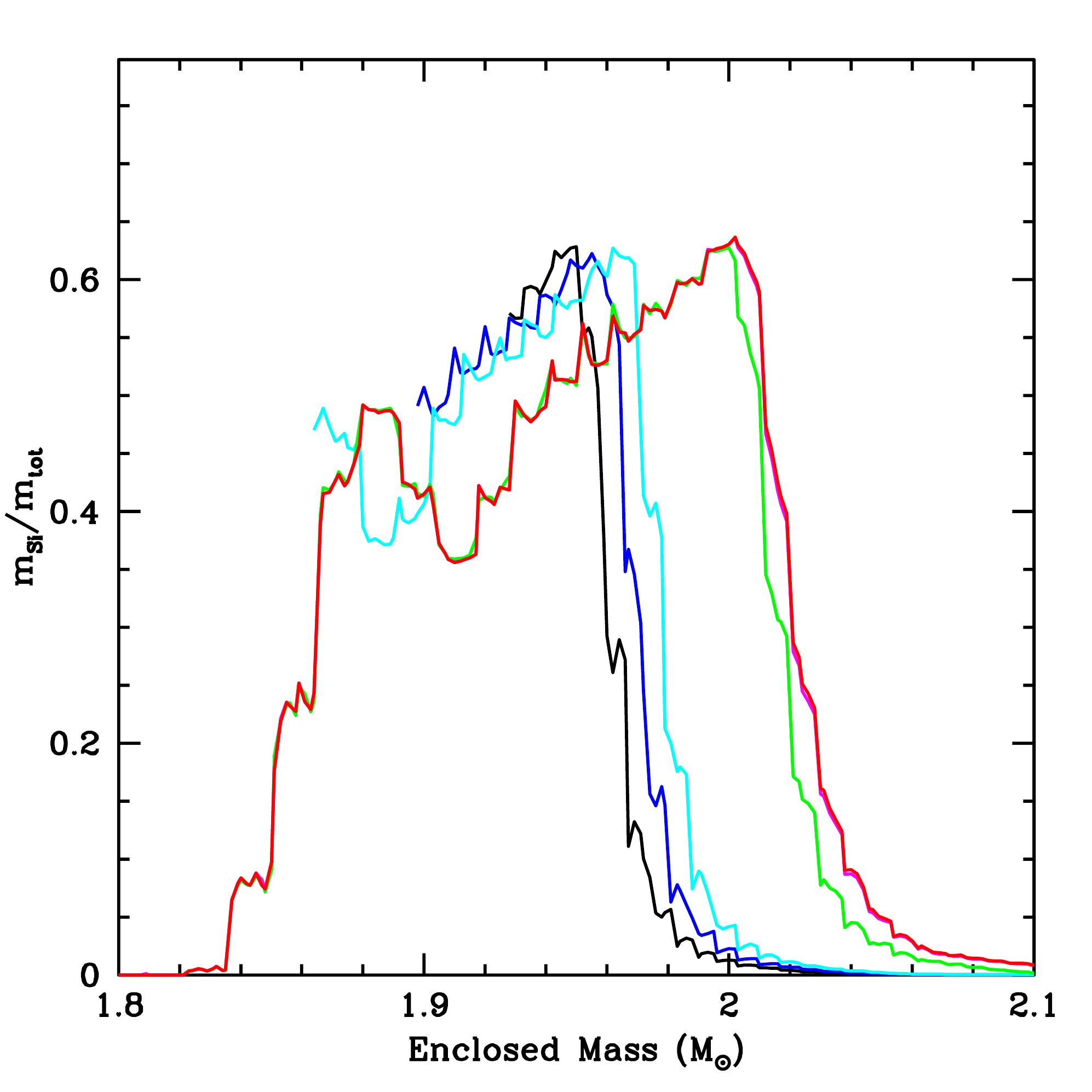}{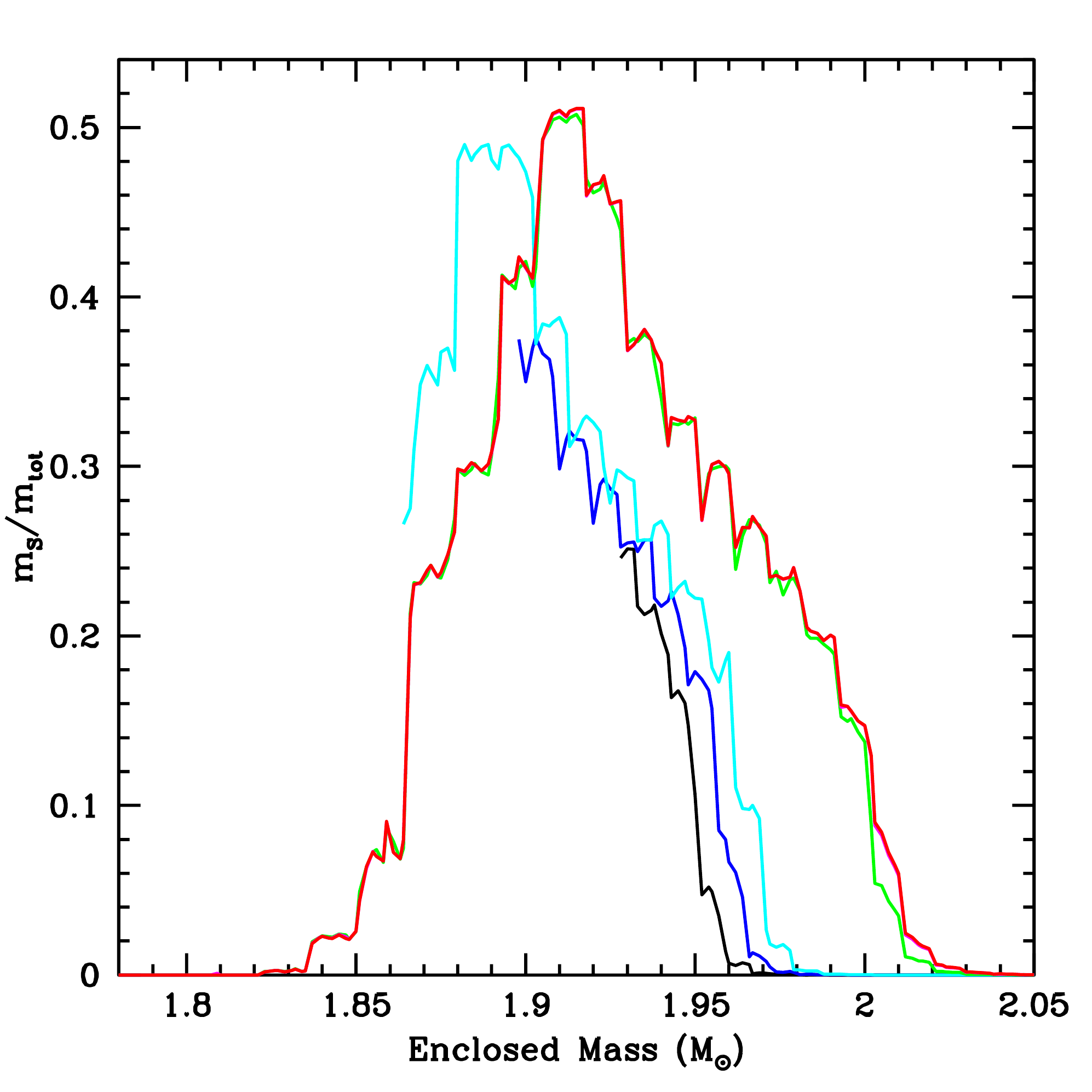}
\plottwo{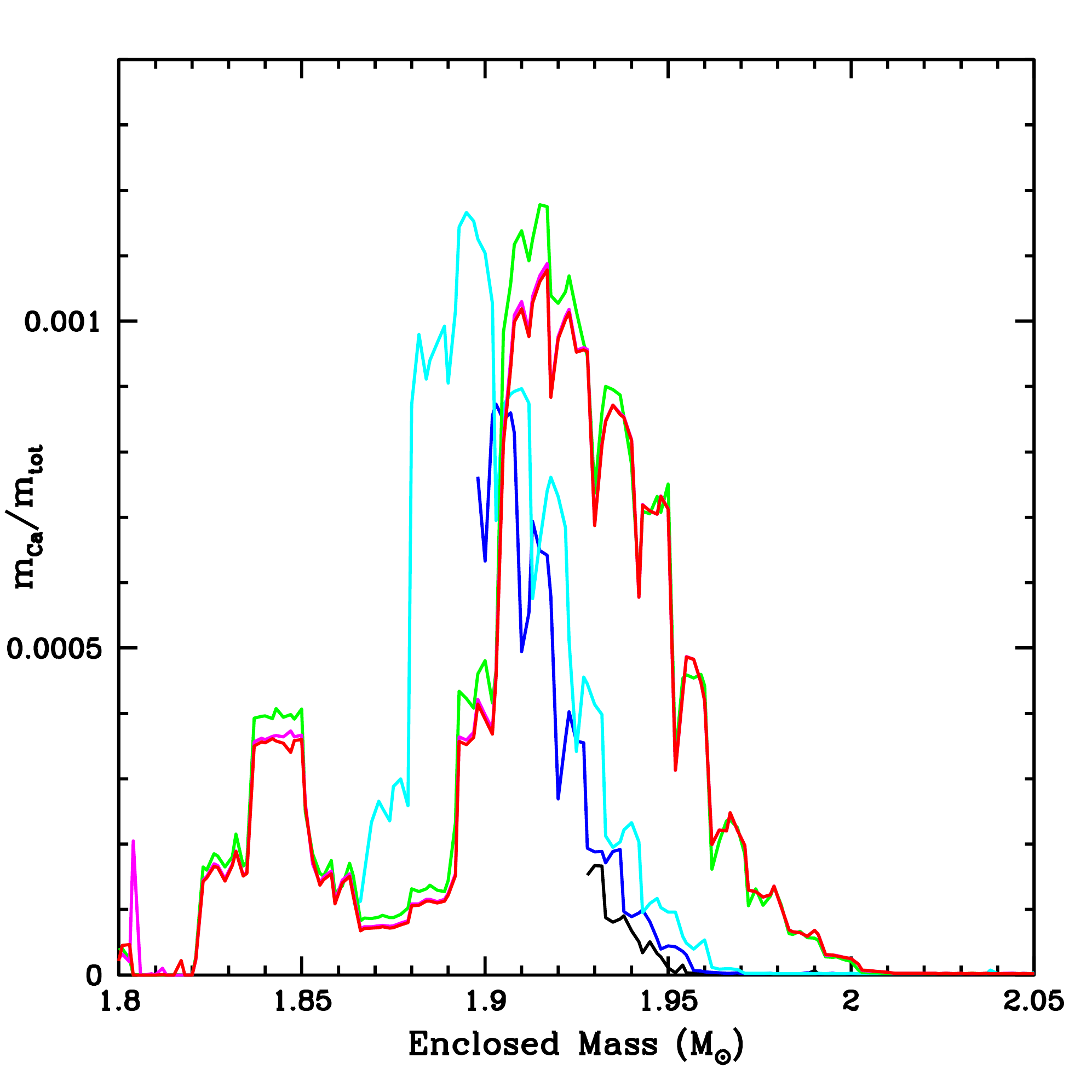}{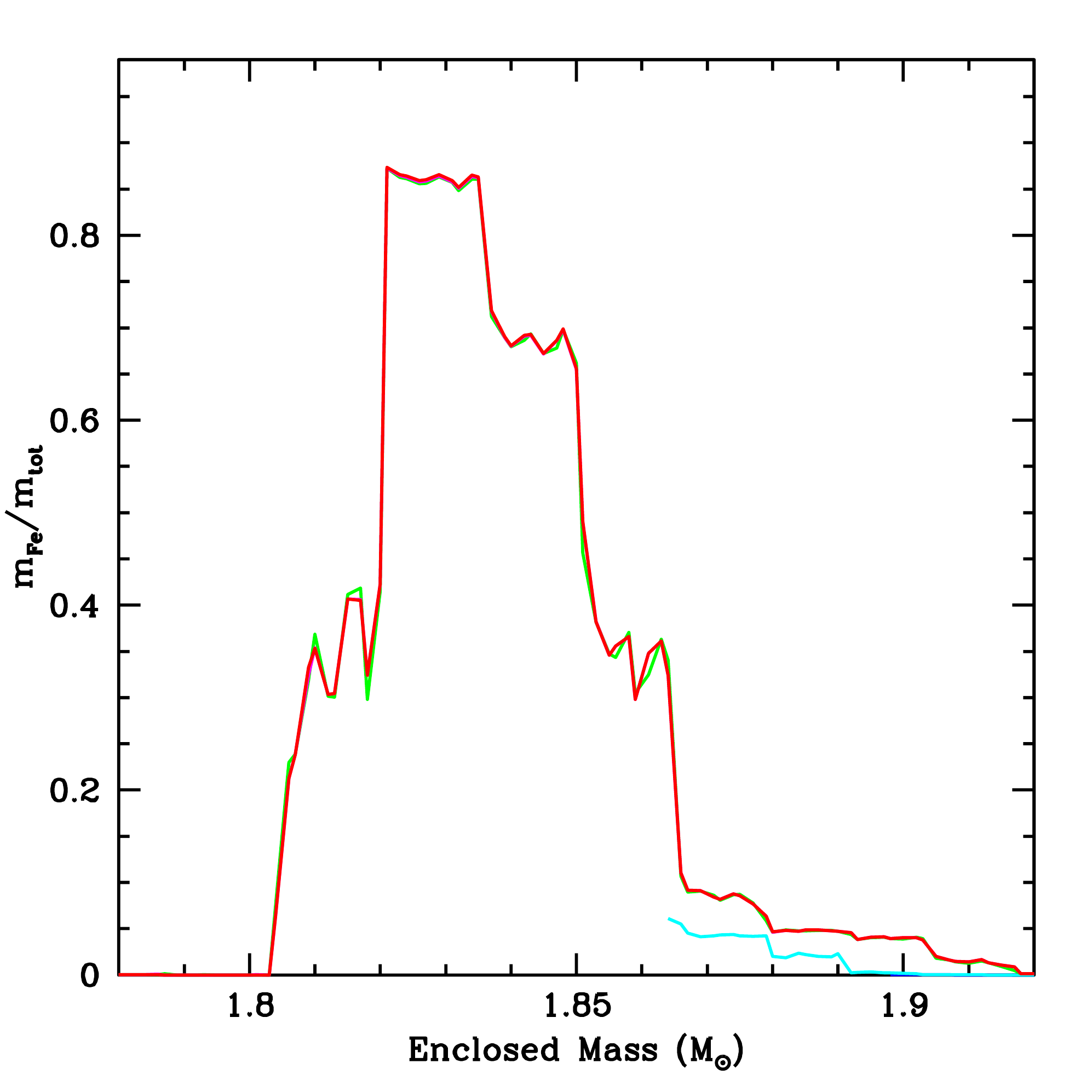}
\epsscale{1.0}

\caption{Magnesium, neon, silicon, sulfur, calcium and iron abundances
  as a function of enclosed mass for the 15\,M$_\odot$ a-series explosions:
  M15aE0.34 (black), M15aE0.58 (blue), M15aE0.82 (cyan), M15aE2.47
  (green), M15aE3.63 (magenta), M15aE4.79 (red).  In the lowest energy
  explosions, considerable material falls back, and those curves start
  at the innermost ejecta mass (below that start, the mass falls back).}

\label{fig:trend}
\end{figure*}

The wide variation in abundances arises from a few basic trends in
explosive nucleosynthesis: fallback and the production and destruction
of elements in a supernova shock.  For very weak explosions, material
is not ejected with sufficient velocity and falls back onto the
neutron star.  The yield of the elements produced near the
proto-neutron star is most affected by fallback.  The iron yield is
extremely sensitive to this fallback.  Figure~\ref{fig:trend} shows
the neon, magnesium, sulfur, calcium, silicon and iron abundances as a
function of enclosed mass for the 15\,M$_\odot$ progenitor with 6
explosion energies ranging from 0.34 to 4.79$\times 10^{51}$\,erg.  
The iron abundance profile (abundance fraction as a function of 
enclosed mass) for the different-energy explosions is fairly similar, 
but the weak explosions simply don't eject this mass.  The silicon, 
sulfur, and calcium profiles are affected by fallback, but also 
by the shock strength.  As the shock energy increases, it induces 
burning further out, producing more of these elements.  Magnesium 
and neon demonstrate the third trend, the destruction of elements 
as they are fused into heavier elements.  The abundance of neon 
initially increases with explosion energy, but then decreases as 
it is destroyed with the highest explosion energies.  The 
combination of destruction and production make it difficult to 
predict trends in the final yields.  The ratio of elements 
can be even more complex, producing the range of abundance ratios.

\begin{figure*}[!hbtp]
\centering
\epsscale{0.85}
\plottwo{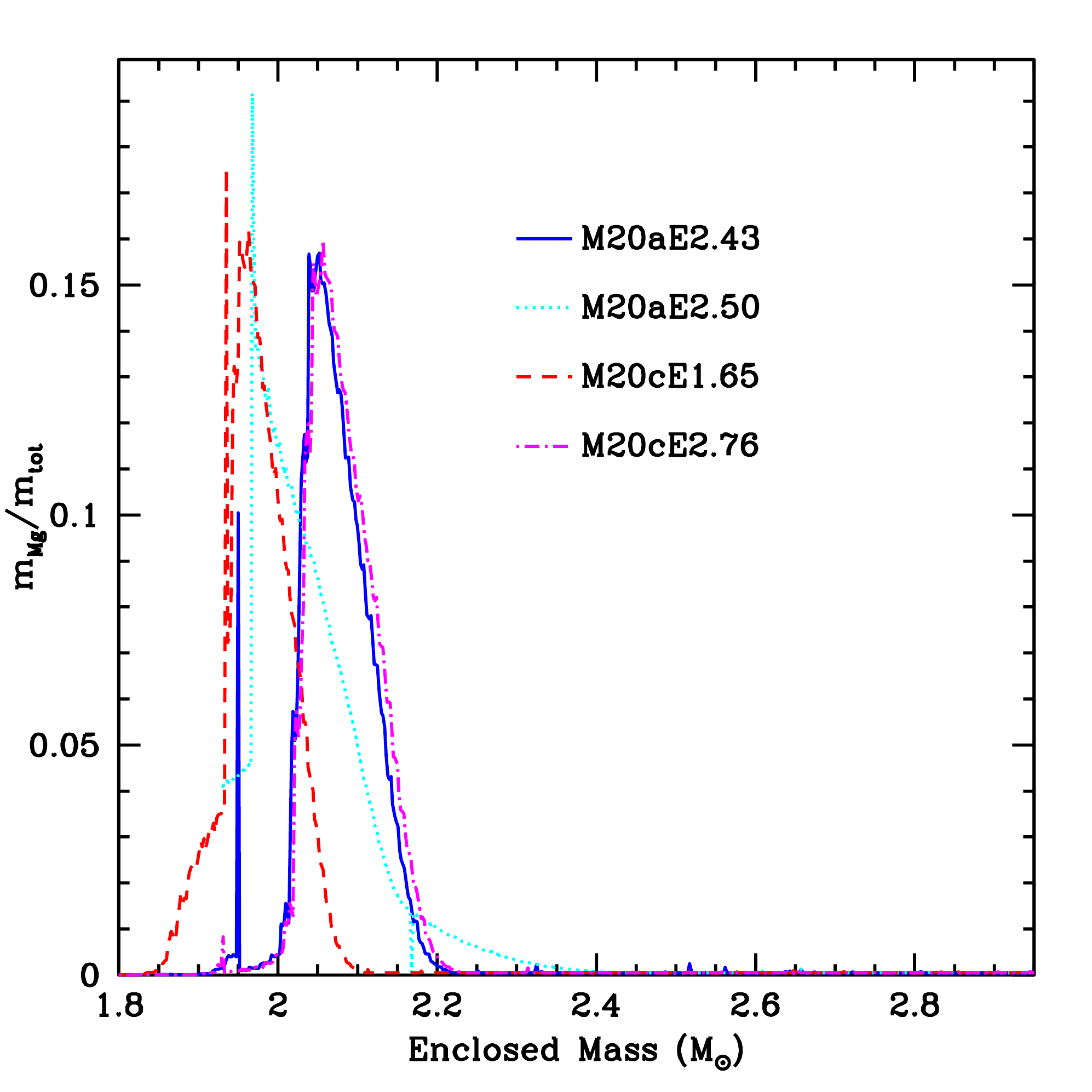}{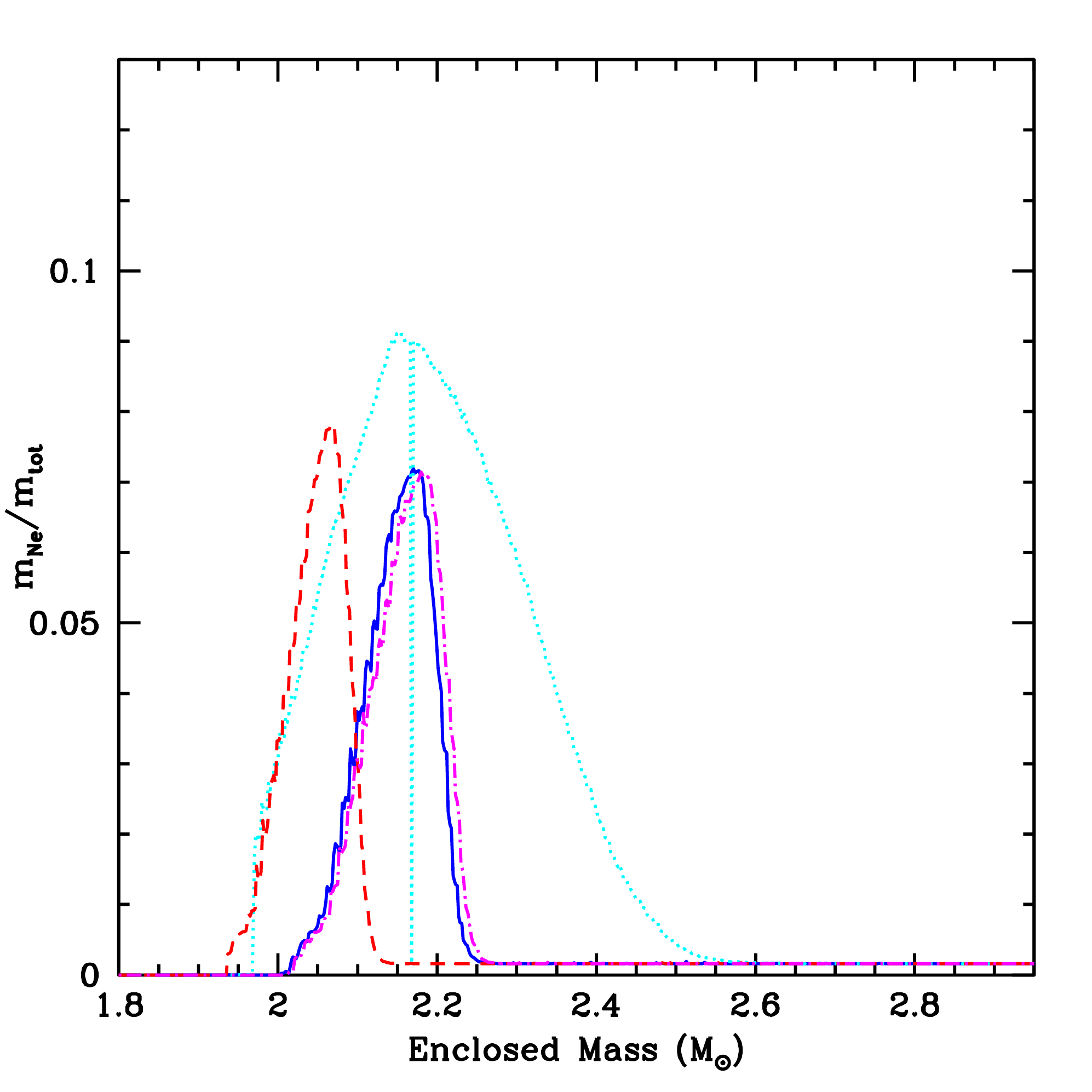}
\plottwo{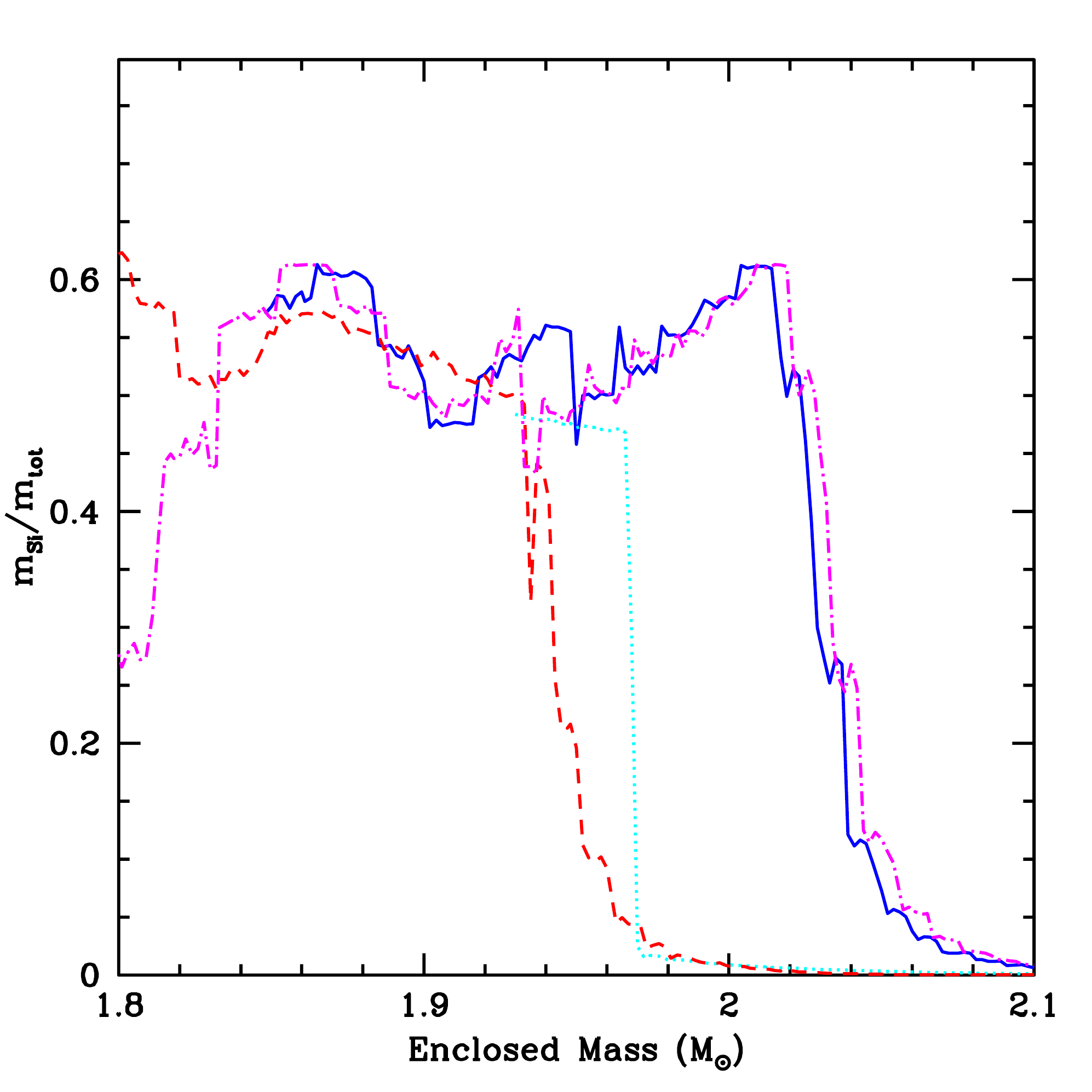}{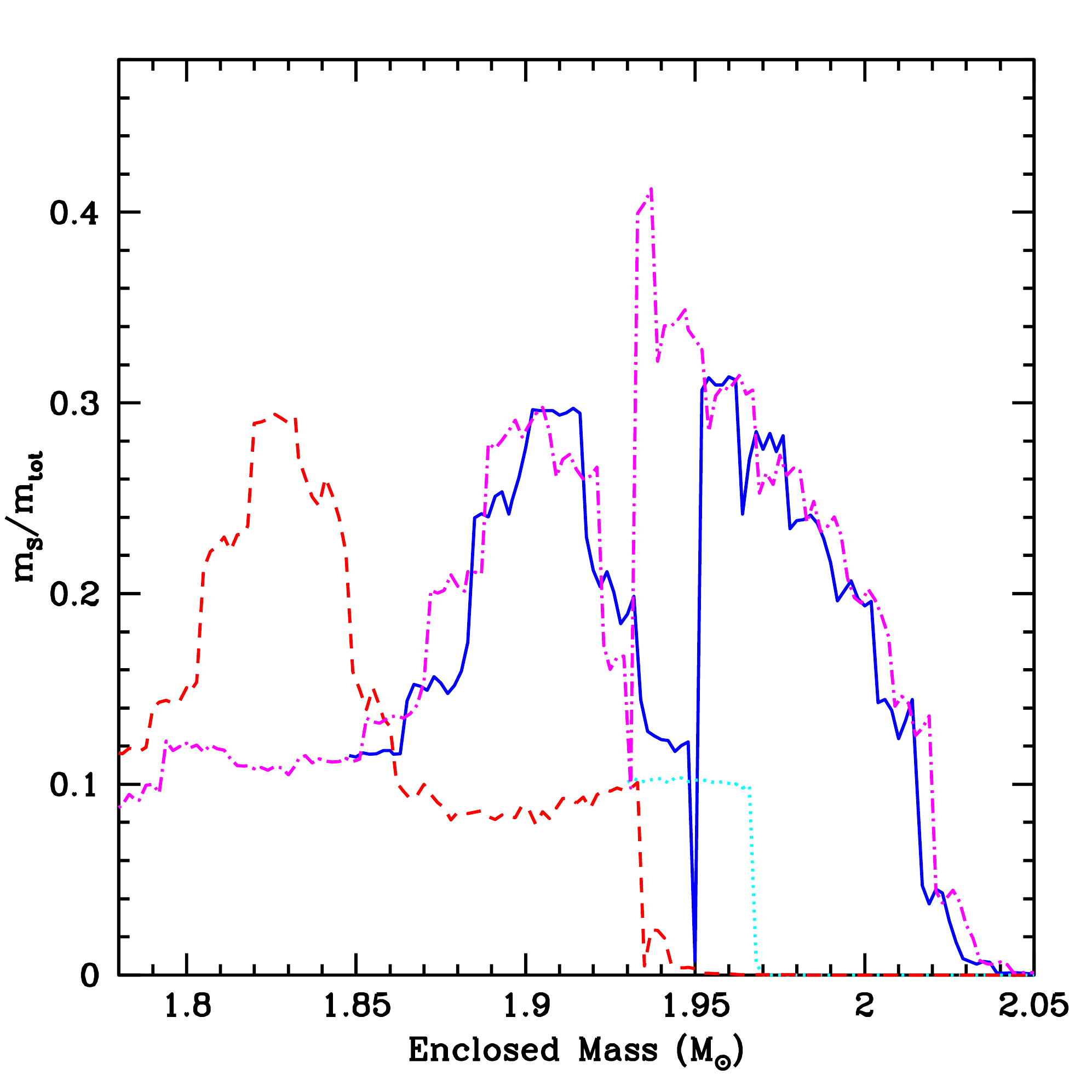}
\plottwo{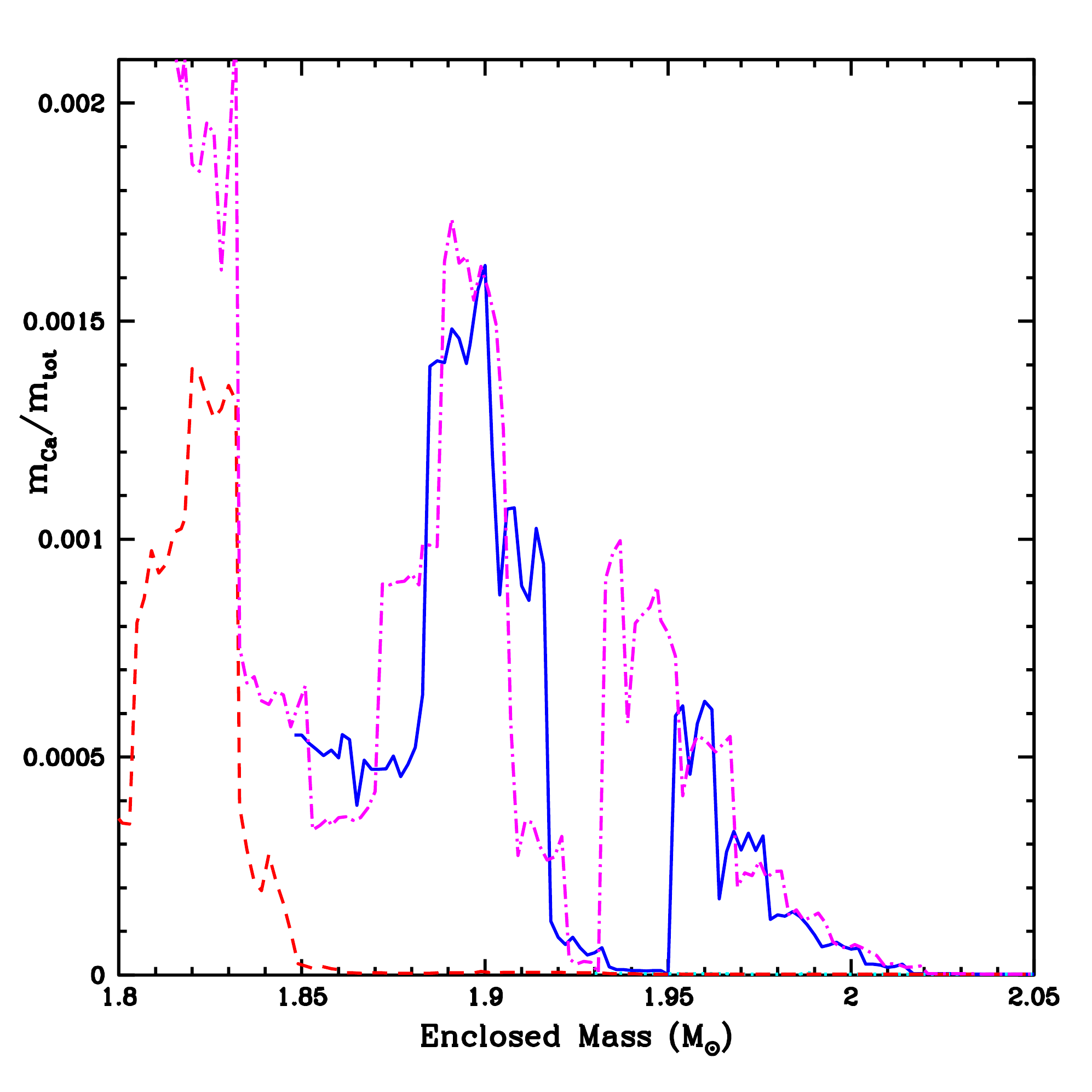}{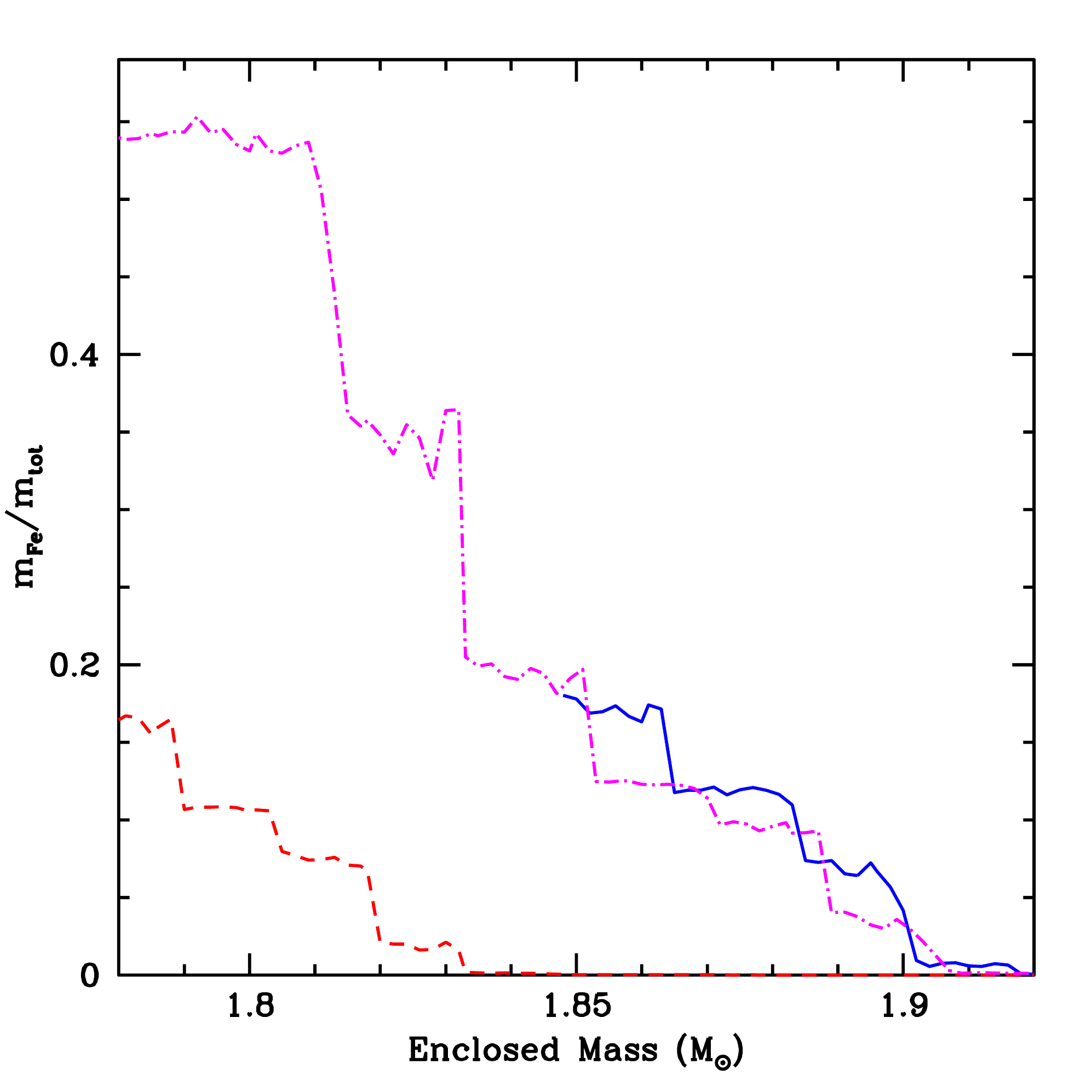}
\epsscale{1.0}

\caption{Magnesium, neon, silicon, sulfur, calcium and iron abundances
  as a function of enclosed mass for four 20\,M$_\odot$ explosions:
  M20aE2.43, M20aE2.50, M20cE1.65, M20cE2.76.  Models M20aE2.43 and
  M20aE2.50 have nearly the same final explosion energy and the same
  total injection energy, but because energy injection occured on
  different time-frames.  Models M20cE1.65 and M20cE2.76 have different 
initial core masses from the a-series models.}

\label{fig:trend2}
\end{figure*}

Additional complexities occur in the nature of the
  explosion.  Figure~\ref{fig:trend2} shows magnesium, neon, silicon,
  sulfur, calcium and iron abundances as a function of enclosed mass
  for four 20\,M$_\odot$ explosions varying the injection time in the
  explosion and the core mass.  Comparing two models with very similar
  injection and final energies, but different injection timescales, we
  see that the yields can vary dramatically.  The longer injection
  timescale can change both the amount of fallback and the strength of
  the shock.  Figure~\ref{fig:trend2} also shows the differences
  between models using slightly different initial core masses.  The
  different core masses reflect the time it takes for convection to
  become strong in the convective engine.  If the growth time of the
  instability is long, more mass will accrete onto the proto-neutron
  star before the convective region robustly distributes energy across
  the convective region.  The profiles, especially at higher mass
  coordinates, are not too different between models with different
  initial core masses.  But the yields can be very different for
  elements, such as iron, that are produced in the innermost zones.
  Understanding the exact nature of the explosion is critical in
  determining the yields.

\section{Conclusions}

Within the paradigm of the convective engine we can produce a wide
range of remnant masses and nucleosynthetic yields.  Even so, some
generic trends exist.  First, even weak explosions are able to eject
most of the star for the 15\,M$_\odot$ star in this study because of
its low binding energy (which is in agreement with most 15\,M$_\odot$
stars).  If an explosion occurs for such progenitors, the compact
remnant is most likely going to be a neutron star.  For more massive
progenitors, late-time energy injection (either through fallback or
magnetar-like activity) is required to make neutron star remnants with
$10^{51}$\,erg explosions.  Without this late energy injection
preventing fallback, more massive progenitors are likely to form black
holes, even if $10^{51}$\,erg explosion occurs.  For the 25\,M$_\odot$
progenitor, an explosion with energy less than $2\times10^{51}$\,erg
without late-time energy injection will produce a black hole.  If we
assume the kick mechanism is produced by asymmetries in the ejecta,
these systems are likely to be the ones that form kicks in black hole
systems~\citep{willems05,fragos09}.

\begin{figure}[!hbtp]
\centering
\plotone{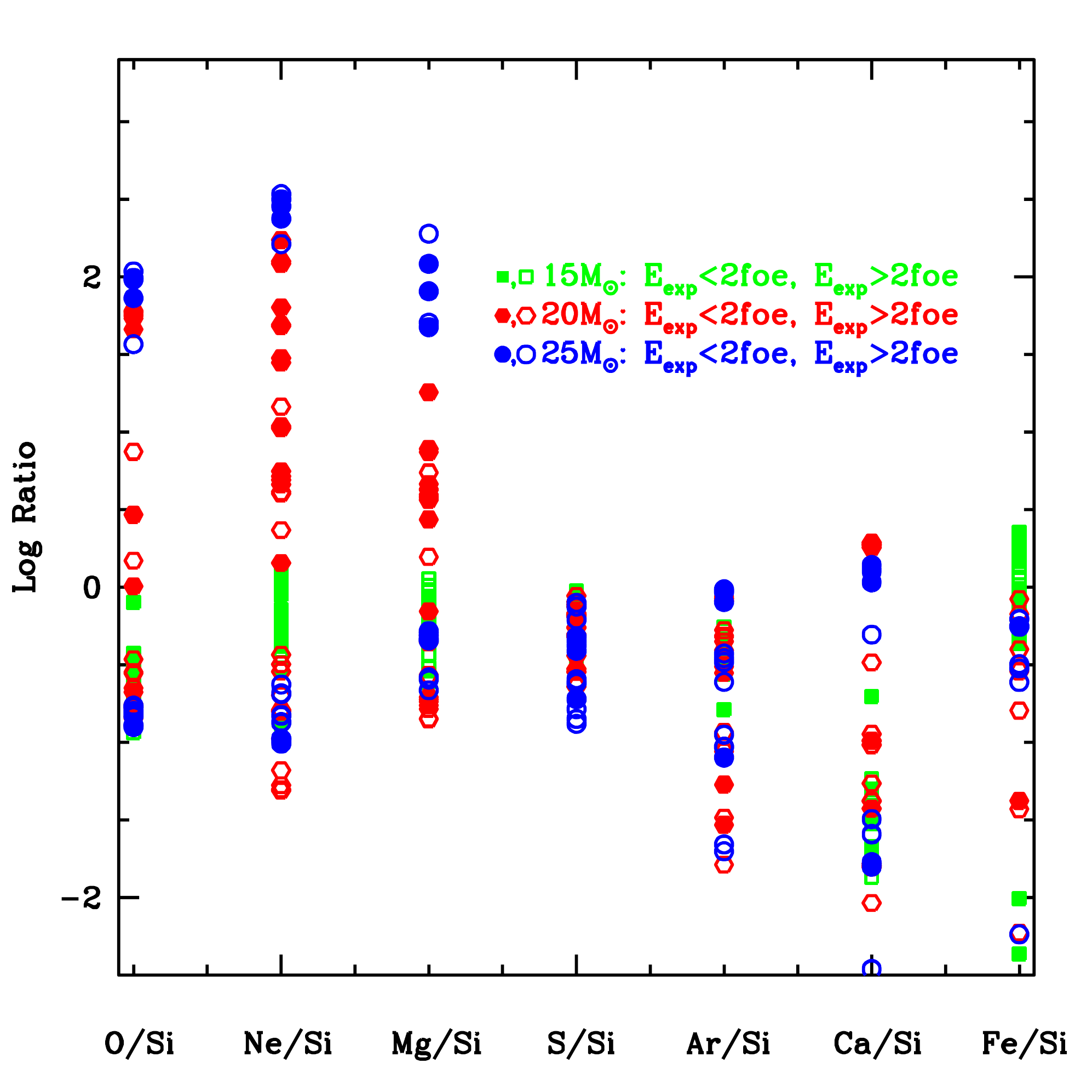}
\caption{Log ratio of a given element to Si with respect to solar
  abundances for a range of elements and all of our models:
  15M$_\odot$ (circle), 20M$_\odot$ (triangle), 25M$_\odot$ (square).
  For the yield ratio, we can get a range of results.  The solid
  symbols refer to explosion energies less than $2\times10^{51}$\,erg, 
open symbols refer to explosion energies above $2\times10^{51}$\,erg.}
\label{fig:abunrat}
\end{figure}

These abundances are inferred in the observations of a number of
supernova remnants~\citep[e.g.,][]{kumar12} and abundance ratios have
been used to place constraints on the supernova progenitor.
Figure~\ref{fig:abunrat} shows the abundance ratio of a few key
elements with respect to silicon: oxygen, neon, magnesium, sulfur,
argon, calcium and iron peak.  Each of our progenitors predict a range
of yields based on the supernova explosion and in many cases, it will
be difficult to discriminate the exact progenitor mass based on the
yields.  There are a few trends: more massive progenitors can produce
more intermediate elements per silicon atom (oxygen, neon, magnesium)
and less massive progenitors produce more iron elements vs. slicon.
As we go up in mass beyond $25\,M_\odot$, stellar winds and mass-loss
through binary interactions (e.g. a common envelope phase) can eject
these intermediate elements.  But if mass-loss is not strong, we
expect these more massive stars to not produce supernovae, ejecting no
heavy elements.  Even these specific trends may be a particular
characteristic of the progenitors we used (from the
KEPLER~\citep{weaver78,woosley02,heger10} code).  The silicon shell is
more massive in the recent TYCHO~\citep{young01} code progenitors,
characteristic of the progenitors we used (from the KEPLER code).  The
silicon shell is more massive in the recent TYCHO code progenitors
produced with modified mixing prescriptions~\cite[e.g.,][]{arnett10}.
More detailed studies of both progenitor and explosion yields will be
required to truly understand the nucleosynthetic yields needed for
discriminating supernova progenitors and in studying galactic chemical
evolution.

With our parameterized study, we have found that the yields depend 
not only on the progenitor and final explosion energy, but on the 
nature of the explosion itself.

{\bf Acknowledgments} This project was funded in part under the
auspices of the U.S. Dept. of Energy, and supported by its contract
W-7405-ENG-36 to Los Alamos National Laboratory.  Simulations at LANL
were performed on HPC resources provided under the Institutional
Computing program.  This research was supported in part by the
National Science Foundation under Grant No. NSF PHY11-25915 and under
LANL LDRD grant 20160173ER.  The Joint Institute for Nuclear
Astrophysics funded workshops that led to this research.  The work of
A.H. was supported by the Australian Research Council ARC FT120100363.
The work of S.S.-H. was supported by NSERC (Natural Sciences and
Engineering Research Council of Canada) and CSA (the Canadian Space
Agency).

\begin{deluxetable*}{lccccccc}
\tablewidth{0pt} 
\tablecaption{Remnant Masses}
\tablehead{\colhead{Model}
& \colhead{M$_{\rm prog}$}
& \colhead{M$_{\rm bounce}$}
& \colhead{M$_{\rm inj}$}
& \colhead{t$_{\rm inj}$\tablenotemark{a}}
& \colhead{E$_{\rm inj}$}
& \colhead{E$_{\rm exp}$}
& \colhead{M$_{\rm remnant}$} \\
\colhead{} 
& \colhead{$(M_\odot)$} 
& \colhead{$(M_\odot)$} 
& \colhead{$(M_\odot)$} 
& \colhead{s}
& \colhead{$10^{51}$\,erg}
& \colhead{$10^{51}$\,erg}
& \colhead{$(M_\odot)$} 
}

\startdata

M15aE0.34  & 15 & 1.30 & 0.3 & 0.1 & 3 & 0.34 & 1.94 \\
M15aE0.54  & 15 & 1.30 & 0.3 & 0.1 & 4 & 0.54 & 1.91 \\
M15aE0.82  & 15 & 1.30 & 0.3 & 0.1 & 5 & 0.82 & 1.88 \\
M15aE2.47  & 15 & 1.30 & 0.3 & 0.1 & 9 & 2.47 & 1.52 \\
M15aE3.63 & 15 & 1.30 & 0.3 & 0.3 & 10 & 3.63 & 1.51 \\
M15aE4.79 & 15 & 1.30 & 0.3 & 0.4 & 20 & 4.79 & 1.50 \\
M15bE0.74 & 15 & 1.30 & 0.02 & 0.4 & 3 & 0.74 & 1.73 \\
M15bE0.92 & 15 & 1.30 & 0.02 & 0.3 & 4 & 0.92 & 1.75 \\
M15bE1.37 & 15 & 1.30 & 0.02 & 0.2 & 5 & 1.37 & 1.80 \\
M15bE1.43 & 15 & 1.30 & 0.02 & 0.1 & 6 & 1.43 & 1.75 \\
M15bE1.48 & 15 & 1.30 & 0.02 & 0.2 & 7 & 1.48 & 1.79 \\
M15bE1.69 & 15 & 1.30 & 0.02 & 0.2 & 10 & 1.69 & 1.52 \\
M15bE2.63 & 15 & 1.30 & 0.02 & 0.2 & 20 & 2.63 & 1.53 \\
M15bE5.08 & 15 & 1.30 & 0.02 & 0.2 & 40 & 5.08 & 1.50 \\
M15bE10.7 & 15 & 1.30 & 0.02 & 0.2 & 80 & $>10.7$ & 1.53 \\
M15cE0.49  & 15 & 1.30 & 0.1 & 0.4 & 3 & 0.49 & 1.89 \\
M15cE0.98  & 15 & 1.30 & 0.1 & 0.4 & 5 & 0.98 & 1.89 \\
M15cE1.79  & 15 & 1.30 & 0.1 & 0.4 & 7 & 1.79 & 1.65 \\
M15cE1.81  & 15 & 1.30 & 0.1 & 0.4 & 8 & 1.81 & 1.63 \\
M15cE1.86 & 15 & 1.30 & 0.1 & 0.3 & 9 & 1.86 & 1.63 \\
M15cE1.90 & 15 & 1.30 & 0.1 & 0.3 & 10 & 1.90 & 1.62 \\
M15cE1.94 & 15 & 1.30 & 0.1 & 0.3 & 12 & 1.94 & 1.61 \\
M15cE2.06 & 15 & 1.30 & 0.1 & 0.3 & 15 & 2.06 & 1.59 \\
M15cE2.24 & 15 & 1.30 & 0.1 & 0.3 & 25 & 2.24 & 1.56 \\
M15cE2.60 & 15 & 1.30 & 0.1 & 0.3 & 45 & 2.60 & 1.52 \\
M15cE3.43 & 15 & 1.30 & 0.1 & 0.3 & 90 & 3.43 & 1.51 \\

\hline

M20aE0.53 & 20 & 1.56 & 0.1 & 0.50 & 4 & 0.53 & 3.40 \\ 
M20aE0.65 & 20 & 1.56 & 0.1 & 0.12 & 4 & 0.65 & 3.03 \\ 
M20aE0.81 & 20 & 1.56 & 0.1 & 0.12 & 7 & 0.81 & 2.70 \\ 
M20aE0.85 & 20 & 1.56 & 0.1 & 0.50 & 7 & 0.85 & 2.62 \\ 
M20aE1.39 & 20 & 1.56 & 0.1 & 0.12 & 10 & 1.39 & 1.93 \\ 
M20aE1.47 & 20 & 1.56 & 0.1 & 0.50 & 10 & 1.47 & 2.23 \\ 
M20aE2.43 & 20 & 1.56 & 0.1 & 0.12 & 20 & 2.43 & 1.86 \\ 
M20aE2.50 & 20 & 1.56 & 0.1 & 0.50 & 20 & 2.50 & 1.93 \\ 
M20aE4.15 & 20 & 1.56 & 0.1 & 0.12 & 50 & 4.15 & 1.85 \\ 
M20bE0.78  & 20 & 1.56 & 0.2 & 0.12 & 5 & 0.78 & 2.85 \\ 
M20bE1.04  & 20 & 1.56 & 0.2 & 0.12 & 6 & 1.04 & 2.47 \\ 
M20bE1.19  & 20 & 1.56 & 0.2 & 0.12 & 8 & 1.19 & 2.28 \\ 
M20bE1.52  & 20 & 1.56 & 0.2 & 0.12 & 10 & 1.52 & 1.97 \\ 
M20bE2.60  & 20 & 1.56 & 0.2 & 0.12 & 25 & 2.60 & 1.90 \\ 
M20bE4.33  & 20 & 1.56 & 0.2 & 0.12 & 50 & 4.33 & 1.87 \\ 
M20cE0.75 & 20 & 1.47 & 0.1 & 0.5 & 6 & 0.75 & 2.76 \\
M20cE0.84 & 20 & 1.47 & 0.1 & 0.5 & 7 & 0.84 & 2.62 \\
M20cE1.00 & 20 & 1.47 & 0.1 & 0.5 & 8 & 1.00 & 2.35 \\
M20cE1.65 & 20 & 1.47 & 0.1 & 0.5 & 10 & 1.65 & 1.78 \\
M20cE2.76 & 20 & 1.47 & 0.1 & 0.5 & 15 & 2.76 & 1.76 \\
M20cE2.85 & 20 & 1.47 & 0.1 & 0.5 & 20 & 2.85 & 1.74 \\
M20cE5.03 & 20 & 1.47 & 0.1 & 0.5 & 50 & 5.03 & 1.74 \\
M20cE8.86 & 20 & 1.47 & 0.1 & 0.5 & 100 & 8.86 & 1.74 \\

\hline

M25aE0.99  & 25 & 1.83 & 0.1 & 0.1 & 5.0 & 0.99 & 4.89 \\ 
M25aE1.57  & 25 & 1.83 & 0.1 & 0.1 & 10 & 1.57 & 3.73 \\ 
M25aE4.73  & 25 & 1.83 & 0.1 & 0.1 & 20 & 4.73 & 2.38 \\ 
M25aE6.17  & 25 & 1.83 & 0.1 & 0.1 & 35 & 6.17 & 2.38 \\ 
M25aE7.42  & 25 & 1.83 & 0.1 & 0.1 & 50 & 7.42 & 2.37 \\ 
M25aE14.8  & 25 & 1.83 & 0.1 & 0.1 & 100 & 14.8 & 2.35 \\ 

M25bE0.76  & 25 & 1.83 & 0.02 & 0.3 & 4.0 & 0.76 & 5.54 \\ 
M25bE1.86  & 25 & 1.83 & 0.02 & 0.5 & 6.0 & 1.86 & 3.52 \\
M25bE1.92  & 25 & 1.83 & 0.02 & 1 & 8.0 & 1.92 & 3.13 \\ 
M25bE8.40  & 25 & 1.83 & 0.02 & 0.28 & 50.0 & 8.40 & 2.38 \\
M25bE9.73  & 25 & 1.83 & 0.02 & 0.69 & 100 & 9.73 & 2.35 \\
M25bE18.4  & 25 & 1.83 & 0.02 & 0.69 & 200 & 18.4 & 2.35 \\
M25d1E3.30 & 25 & 1.83 & 0.02 & 0.7,10-30\tablenotemark{b} & 25 & 3.30 & 2.35 \\
M25d1E4.72 & 25 & 1.83 & 0.02 & 0.7,10-30\tablenotemark{b} & 50 & 4.72 & 2.35 \\
M25d1E7.08 & 25 & 1.83 & 0.02 & 0.7,10-30\tablenotemark{b} & 100 & 7.08 & 2.35 \\
M25d2E2.53 & 25 & 1.83 & 0.02 & 0.7,100-300\tablenotemark{b} & 20 & 2.53 & 2.35 \\
M25d2E2.64 & 25 & 1.83 & 0.02 & 0.7,100-300\tablenotemark{b} & 35 & 2.64 & 2.35 \\
M25d2E2.78 & 25 & 1.83 & 0.02 & 0.7,100-300\tablenotemark{b} & 50 & 2.78 & 2.35 \\
M25d2E3.07 & 25 & 1.83 & 0.02 & 0.7,100-300\tablenotemark{b} & 100 & 3.07 & 1.83 \\
M25d3E0.89 & 25 & 1.83 & 0.02 & 0.7,1000-3000\tablenotemark{b} & 7 & 0.89 & 4.66 \\
M25d3E0.92 & 25 & 1.83 & 0.02 & 0.7,1000-3000\tablenotemark{b} & 8 & 0.92 & 1.84 \\
M25d3E1.04 & 25 & 1.83 & 0.02 & 0.7,1000-3000\tablenotemark{b} & 10 & 1.04 & 1.84 \\
M25d3E1.20 & 25 & 1.83 & 0.02 & 0.7,1000-3000\tablenotemark{b} & 50 & 1.20 & 1.84 \\
M25d3E1.52 & 25 & 1.83 & 0.02 & 0.7,1000-3000\tablenotemark{b} & 100 & 1.52 & 1.83 \\
\enddata
\tablenotetext{a}{The core is defined by the post-merger material whose 
density is above $10^{14} {\rm g cm^{-3}}$ at the end of the calculation.}
\tablenotetext{b}{The core is defined by the post-merger material whose 
density is above $10^{14} {\rm g cm^{-3}}$ at the end of the calculation.}
\label{tab:runs}

\end{deluxetable*}

\begin{deluxetable*}{lcccccccc}
\tablewidth{0pt} 
\tablecaption{Explosive Yields}
\tablehead{\colhead{Model}
& \colhead{M$_O (M_\odot)$}
& \colhead{M$_{Ne} (M_\odot)$}
& \colhead{M$_{Mg} (M_\odot)$}
& \colhead{M$_{Si} (M_\odot)$}
& \colhead{M$_{S} (M_\odot)$}
& \colhead{M$_{Ar} (M_\odot)$}
& \colhead{M$_{Ca} (M_\odot)$}
& \colhead{M$_{Fe} (M_\odot)$}
}

\startdata

M15aE0.34 & 0.29 & 0.064 & 0.013 & 0.022 & 0.0051 & 0.00021 & $4.4\times10^{-5}$ & $5.5\times10^{-8}$ \\
M15aE0.58 & 0.25 & 0.11 & 0.027 & 0.043 & 0.017 & 0.00081 & $6.4\times10^{-5}$ & $2.9\times10^{-6}$ \\
M15aE0.82 & 0.25 & 0.11 & 0.041 & 0.073 & 0.040 & 0.0033 & $6.4\times10^{-5}$ & 0.0010 \\
M15aE2.47 & 0.25 & 0.098 & 0.057 & 0.083 & 0.045 & 0.0037 & 0.00015 & 0.16 \\
M15aE3.63 & 0.27 & 0.086 & 0.083 & 0.089 & 0.046 & 0.0037 & 0.00032 & 0.29 \\
M15aE4.79 & 0.27 & 0.0085 & 0.087 & 0.093 & 0.046 & 0.0037 & 0.00029 & 0.30 \\
M15bE0.74 & 0.28 & 0.11 & 0.054 & 0.072 & 0.040 & 0.0033 & 0.00012 & 0.12 \\
M15bE0.92 & 0.27 & 0.11 & 0.054 & 0.071 & 0.040 & 0.0033 & 0.00012 & 0.10 \\
M15bE1.37 & 0.25 & 0.11 & 0.041 & 0.073 & 0.040 & 0.0034 & 0.00011 & 0.054 \\
M15bE1.43 & 0.25 & 0.11 & 0.043 & 0.074 & 0.041 & 0.0034 & 0.00013 & 0.098 \\
M15bE1.48 & 0.25 & 0.10 & 0.045 & 0.073 & 0.041 & 0.0034 & 0.00011 & 0.065 \\
M15bE1.69 & 0.26 & 0.010 & 0.0052 & 0.072 & 0.040 & 0.0034 & 0.00039 & 0.27 \\
M15bE2.63 & 0.26 & 0.0088 & 0.075 & 0.087 & 0.046 & 0.0037 & 0.00034 & 0.27 \\
M15bE5.08 & 0.24 & 0.081 & 0.077 & 0.13 & 0.076 & 0.0069 & 0.00043 & 0.35 \\
M15vE10.7 & 0.26 & 0.10 & 0.052 & 0.11 & 0.051 & 0.0037 & 0.00040 & 0.27 \\
M15cE0.49 & 0.26 & 0.11 & 0.023 & 0.047 & 0.023 & 0.0016 & $8.0\times10^{-5}$ & $6.4\times10^{-5}$ \\
M15cE0.98 & 0.25 & 0.11 & 0.034 & 0.054 & 0.030 & 0.0023 & $8.9\times10^{-5}$ & 0.00040 \\
M15cE1.79 & 0.27 & 0.023 & 0.013 & 0.084 & 0.047 & 0.0034 & 0.0015 & 0.29 \\
M15cE1.81 & 0.10 & 0.047 & 0.047 & 0.075 & 0.041 & 0.0035 & 0.00021 & 0.19 \\
M15cE1.86 & 0.25 & 0.10 & 0.048 & 0.075 & 0.041 & 0.0035 & 0.00021 & 0.19 \\
M15cE1.90 & 0.25 & 0.10 & 0.051 & 0.075 & 0.041 & 0.0035 & 0.00024 & 0.19 \\
M15cE1.94 & 0.25 & 0.10 & 0.051 & 0.075 & 0.041 & 0.0035 & 0.00024 & 0.20 \\
M15cE2.06 & 0.25 & 0.099 & 0.054 & 0.075 & 0.042 & 0.0035 & 0.00027 & 0.22 \\
M15cE2.24 & 0.25 & 0.094 & 0.059 & 0.075 & 0.042 & 0.0035 & 0.00031 & 0.24 \\
M15cE2.60 & 0.25 & 0.090 & 0.066 & 0.075 & 0.042 & 0.0035 & 0.00034 & 0.27 \\
M15cE3.43 & 0.25 & 0.089 & 0.079 & 0.078 & 0.042 & 0.0035 & 0.00033 & 0.29 \\

\hline

M20aE0.53  & 0.29 & 0.0031 & 0.0011 & 0.00032 & 0.00015 & $3.1\times10^{-5}$ & $5.7\times10^{-5}$ & 0. \\
M20aE0.65  & 0.34 & 0.0085 & 0.0012 & 0.00037 & 0.00017 & $3.6\times10^{-5}$ & $6.4\times10^{-5}$ & 0. \\
M20aE0.81  & 0.40 & 0.041 & 0.0016 & 0.00041 & 0.00019 & $4.1\times10^{-5}$ & $7.1\times10^{-5}$ & 0.0 \\
M20aE0.85  & 0.40 & 0.056 & 0.0016 & 0.00045 & 0.00020 & $4.1\times10^{-5}$ & $7.3\times10^{-5}$ & 0. \\
M20aE1.39  & 0.40 & 0.073 & 0.015 & 0.025 & 0.0048 & $8.5\times10^{-5}$ & $8.4\times10^{-5}$ & $3.8\times10^{-7}$ \\
M20aE1.47  & 0.33 & 0.0052 & 0.0016 & 0.00042 & 0.00020 & $4.6\times10^{-5}$ & $7.4\times10^{-5}$ & 0. \\
M20aE2.43  & 0.37 & 0.044 & 0.018 & 0.10 & 0.035 & 0.0042 & 0.00016 & 0.0066 \\
M20aE2.50  & 0.50 & 0.070 & 0.037 & 0.092 & 0.018 & 0.00095 & $7.8\times10^{-5}$ & $2.5\times10^{-5}$ \\
M20aE4.15  & 0.36 & 0.044 & 0.022 & 0.13 & 0.068 & 0.0072 & 0.0011 & 0.062 \\

M20bE0.78  & 0.37 & 0.024 & 0.0014 & 0.00039 & 0.0018 & $4.3\times10^{-5}$ & $7.4\times10^{-5}$ & 0.  \\
M20bE1.04  & 0.41 & 0.11 & 0.00030 & 0.00043 & 0.00020 & $4.3\times10^{-5}$ & $7.4\times10^{-5}$ & 0. \\
M20bE1.19 & 0.40 & 0.16 & 0.0073 & 0.00045 & 0.00021 & $4.4\times10^{-5}$ & $7.7\times10^{-5}$ & 0. \\
M20bE1.52 & 0.41 & 0.094 & 0.021 & 0.087 & 0.0017 & $5.5\times10^{-5}$ & $8.1\times10^{-5}$ & $1.9\times10^{-7}$ \\
M20bE2.60 & 0.40 & 0.059 & 0.022 & 0.088 & 0.030 & 0.0032 & 0.00013 & 0.000089 \\
M20bE4.33  & 0.37 & 0.048 & 0.023 & 0.14 & 0.073 & 0.0075 & 0.00071 & 0.039 \\
M20cE0.75 & 0.38 & 0.024 & 0.0014 & 0.00040 & 0.00018 & $4.0 \times 10^{-5}$ & $7.0 \times 10^{-5}$ & 0. \\
M20cE0.84 & 0.40 & 0.043 & 0.0016 & 0.00042 & 0.00019 & $4.1 \times 10^{-5}$ & $7.3 \times 10^{-5}$ & 0. \\
M20cE1.00 & 0.42 & 0.11 & 0.0029 & 0.00043 & 0.00020 & $4.3 \times 10^{-5}$ & $7.4 \times 10^{-5}$ & 0. \\
M20cE1.65 & 0.33 & 0.022 & 0.016 & 0.099 & 0.024 & 0.0032 & 0.00014 & 0.0070 \\
M20cE2.76 & 0.31 & 0.018 & 0.017 & 0.13 & 0.050 & 0.0061 & 0.0012 & 0.064 \\
M20cE2.85 & 0.29 & 0.014 & 0.016 & 0.13 & 0.047 & 0.0058 & 0.0012 & 0.085 \\
M20cE5.03 & 0.30 & 0.014 & 0.019 & 0.13 & 0.071 & 0.0074 & 0.0013 & 0.15 \\
M20cE8.86 & 0.30 & 0.015 & 0.023 & 0.15 & 0.088 & 0.0090 & 0.00056 & 0.21 \\

\hline

M25bE1.86  & 0.68 & 0.28 & 0.018 & 0.00039 & 0.00011 & $5.2\times10^{-5}$ & $5.7\times10^{-5}$ & 0.0 \\
M25bE1.92  & 0.88 & 0.38 & 0.083 & 0.00076 & 0.00020 & $7.1\times10^{-5}$ & $7.5\times10^{-5}$ & $4.0\times10^{-9}$ \\
M25bE8.40 & 1.05 & 0.13 & 0.19 & 0.41 & 0.037 & 0.00096 & $9.9\times10^{-5}$ & $3.2\times10^{-6}$ \\
M25bE9.73 & 1.03 & 0.12 & 0.19 & 0.44 & 0.042 & 0.0011  & 0.00010 & $1.0\times10^{-5}$ \\
M25bE18.4 & 1.04 & 0.12 & 0.19 & 0.43 & 0.042 & 0.0011 & 0.00010 & $9.8\times10^{-6}$ \\

M25d1E3.30 & 0.99 & 0.11 & 0.17 & 0.37 & 0.061 & 0.0040 & 0.00012 & 0.0036 \\
M25d1E4.72 & 0.99 & 0.11 & 0.17 & 0.37 & 0.061 & 0.0040 & 0.00012 & 0.0036 \\
M25d1E7.08 & 0.99 & 0.11 & 0.17 & 0.37 & 0.061 & 0.0040 & 0.00012 & 0.0036 \\
M25d2E2.53 & 0.99 & 0.11 & 0.17 & 0.37 & 0.061 & 0.0040 & 0.00012 & 0.0036 \\
M25d2E2.64 & 0.99 & 0.11 & 0.17 & 0.37 & 0.061 & 0.0040 & 0.00012 & 0.0036 \\
M25d2E2.78 & 0.99 & 0.11 & 0.17 & 0.37 & 0.061 & 0.0040 & 0.00012 & 0.0036 \\
M25d2E3.07 & 0.84 & 0.090 & 0.17 & 0.41 & 0.072 & 0.0054 & 0.00063 & 0.43 \\
M25d3E0.89 & 0.77 & 0.29 & 0.021 & 0.00049 & 0.00016 & $5.5\times10^{-5}$ & $6.2\times10^{-5}$ & 0.0 \\
M25d3E0.92 & 0.73 & 0.078 & 0.15 & 0.37 & 0.048 & 0.0035 & 0.00055 & 0.35 \\
M25d3E1.04 & 0.74 & 0.081 & 0.15 & 0.37 & 0.048 & 0.0034 & 0.00053 & 0.35 \\
M25d3E1.20 & 0.73 & 0.078 & 0.15 & 0.37 & 0.048 & 0.0035 & 0.00055 & 0.35 \\
M25d3E1.52 & 0.74 & 0.079 & 0.15 & 0.37 & 0.048 & 0.0034 & 0.00057 & 0.35 \\
\enddata
\tablenotetext{a}{The core is defined by the post-merger material whose 
density is above $10^{14} {\rm g cm^{-3}}$ at the end of the calculation.}
\tablenotetext{b}{The core is defined by the post-merger material whose 
density is above $10^{14} {\rm g cm^{-3}}$ at the end of the calculation.}

\label{tab:yields}

\end{deluxetable*}

\end{document}